\documentclass[aps,pre,showpacs]{revtex4}
\usepackage{color}
\usepackage{colordvi}
\usepackage{epsfig}
\usepackage{bm,amssymb}
\usepackage{mathtools}

\begin{document}

\title{
Revealing the Micro-Structure 
of the Giant Component \\
in Random Graph Ensembles
}

\author{Ido Tishby}

\affiliation{Racah Institute of Physics, 
The Hebrew University, Jerusalem, 91904, Israel}

\author{Ofer Biham}

\affiliation{Racah Institute of Physics, 
The Hebrew University, Jerusalem, 91904, Israel}

\author{Eytan Katzav}

\affiliation{Racah Institute of Physics, 
The Hebrew University, Jerusalem, 91904, Israel}

\author{Reimer K\"uhn}

\affiliation{Mathematics Department, King's College London, 
Strand, London WC2R 2LS, UK}

\begin{abstract}

The micro-structure of the giant component 
of the Erd{\H o}s-R\'enyi network and other configuration model networks is
analyzed using generating function methods. 
While configuration model networks are uncorrelated, 
the giant component 
exhibits a degree distribution which is different
from the overall degree distribution of the network 
and includes degree-degree correlations of all orders.
We present exact analytical results for the
degree distributions as well as higher order degree-degree correlations
on the giant components of configuration model networks.
We show that the degree-degree correlations are essential for the integrity
of the giant component, in the sense that the degree distribution alone
cannot guarantee that it will consist of a single connected component.
To demonstrate the importance and broad applicability of these results,
we apply them to the study of the distribution of shortest path lengths 
on the giant component, percolation on the giant component and the
spectra of sparse matrices defined on the giant component.
We show that by using the degree distribution on the giant component,
one obtains high quality results for these properties,
which can be further improved by taking the degree-degree correlations into account. 
This suggests that many existing methods, 
currently used for the analysis of the whole network, 
can be adapted in a straightforward fashion
to yield results conditioned on the giant component.

\end{abstract}

\maketitle

\section{Introduction}

There is a broad range of phenomena in the natural sciences and engineering
as well as in the economic and social sciences, which can be usefully
described in terms of network models.
This realization has stimulated increasing interest during the past
two decades in the study of the structure of random graphs and complex networks, 
and in the dynamics of processes which take place on them 
\cite{Albert2002,Dorogovtsev2003,Dorogovtsev2008,Newman2010,Barrat2012,Hofstad2013}.
One of the central lines of inquiry since Erd\H{o}s and 
R\'enyi's seminal study of the evolution of random graphs 
\cite{Erdos1959,Erdos1960,Erdos1961} 
has been concerned with the existence, under suitable conditions,
of a  giant component, which occupies 
a finite, non-zero fraction of the graph in the thermodynamic 
limit of infinite system size. 
Critical parameters for the 
emergence of a giant component in the thermodynamic limit of 
Erd\H{o}s-R\'enyi (ER) networks were identified and the 
asymptotic fraction occupied by the giant component was determined 
\cite{Erdos1960,Bollobas1984}. 
For configuration model networks, 
i.e. networks that are maximally random subject to a given degree 
sequence, those problems were solved by Molloy and Reed 
\cite{Molloy1995,Molloy1998}. 
These authors also 
established a so-called duality relation according to which the 
degree distribution, when restricted to nodes which reside on finite 
components of a configuration model network, is simply related 
to the degree distribution of the whole network 
\cite{Molloy1998}. 
This property, which 
%appears to have been 
was known before for ER networks 
\cite{Bollobas1984,Bollobas2001}, 
has since been generalized also 
to a class of heterogeneous 
{\it canonical} random graph models 
with broad distributions of expected degrees 
\cite{Bollobas2007,Janson2010}.
Curiously, with the single exception of a study on large 
deviation properties of ER networks by Engel et al. 
\cite{Engel2004}, 
we have not come across corresponding 
statements concerning degree distributions when restricted 
to the giant component of a random graph ensemble.
Clearly, the knowledge of degree distributions and
degree-degree correlations restricted to the giant component of a network would 
be very useful when investigating dynamical processes on complex networks.
It would help to obtain results pertaining only to 
the giant component of such systems, without the contributions from 
finite components which often amount to trivial contaminations or (unwanted) 
distortions of results. 
Examples that come to mind are localization 
phenomena in sparse matrix spectra 
\cite{Biroli1999,Kuhn2008} 
(where finite components of a random graph 
support eigenvectors that are trivially localized),  
properties of random walks 
\cite{Sood2005,Bacco2015,Tishby2016,Tishby2017} 
(where a random walker chosen 
to start a walk on one of the finite components will never be able to explore 
an appreciable fraction of the entire network), 
or the spread of diseases or cascading failures 
\cite{Watts2002,Newman2002,Newman2002b,Karrer2010,Rogers2015,Satorras2015} 
(where an initial failure or initial infection occurring on a finite 
component will never lead to a global system failure or the outbreak of 
an epidemic).  
Component-size distributions in the percolation problem 
on complex networks 
\cite{Callaway2000,Rogers2015} 
will likewise contain a 
component originating from clusters that were finite, {\it before} 
nodes or edges were randomly removed from the network. 
Finally, the distribution of shortest path lengths between pairs of nodes in a 
network 
\cite{Blondel2007,Katzav2015,Nitzan2016,Melnik2016} 
contain contributions 
from pairs of nodes on different components whose distance is, by 
convention, infinite.
To eliminate such unwanted contributions, numerical studies using
message passing algorithms, or straightforward simulation methods, 
are often performed directly on the largest component of a given random 
network. 
The results may be  difficult to compare with theoretical 
results if the latter do not eliminate finite component contributions 
(or suppress them by taking the density of links to be sufficiently 
large to make them effectively negligible).

It is the purpose of the present contribution to explore the 
micro-structure of the giant component as well as the finite components 
of random graphs in the configuration model class. 
More specifically, we use generating functions 
\cite{Newman2001} 
and their probabilistic interpretation to 
obtain degree distributions {\it conditioned} on both the giant and 
finite components appearing in networks in the configuration model 
class, as well as degree-degree correlations of all orders in these 
networks. 
The key assumption underlying the use of the generating 
function method is that configuration model networks 
(with finite mean degree), which are only locally tree-like, 
are in fact probabilistically well approximated by trees in the 
limit of large system sizes. The underlying reason is that any 
correlations between neighbours of a given node which are 
generated by the existence of loops become arbitrarily small, 
as the typical lengths of such loops diverge like $\log(N)$ with 
the system size, $N$.

The following are among our key results: 
(i) the giant components of ER networks 
and of configuration model networks
exhibit degree-degree correlations of all orders, and 
are thus not in the configuration model class; 
(ii) degree-degree correlations of all orders need to be taken into account in order to 
verify that {\it conditioned} on the giant component of a random 
network in the configuration model class --- there are indeed no finite 
trees of any size; 
(iii) for finite components of ER and configuration 
model networks one has a duality relation linking the degree distribution 
restricted to the finite components to that of a different sub-percolating 
configuration model via renormalization of the original degree distribution 
over the whole network; 
(iv) we provide examples demonstrating the quality of results that can be obtained 
for properties of the giant component when neglecting degree-degree 
correlations, and improvements that can be made when taking degree-degree 
correlations into account. 

The paper is organized as follows. 
In Sec. \ref{Sec:CM} we present the configuration model
network ensemble.
In Sec. \ref{Sec:GC} 
we recall the generating function approach 
\cite{Newman2001}, 
which allows to 
establish the existence of a giant component in configuration 
model networks, and to compute the asymptotic fraction $g$ of nodes that 
belong to the giant component. 
Concentrating on the probabilistic 
content of individual contributions to the expression for $g$ in that approach 
allows us to extract the degree distributions conditioned on nodes belonging 
to either the giant component or to any of the finite components. 
For the finite components we recover the duality relation obtained 
by Molloy and Reed 
\cite{Molloy1998}. 
Using iterated versions of one 
of the generating functions then allows us in Sec.
\ref{Sec:JointGC} 
to obtain 
joint degree distributions (and thereby degree-degree correlations) of all orders for 
nodes surrounding a given node, conditioned on the central 
node belonging to either the giant component or to one of the finite components.
In Sec.  \ref{Sec:assort} we derive an analytical expression for the assortativity
coefficient on the giant component of a configuration model network with any given
degree distribution.
In Sec. \ref{Sec:examp} we apply the results to several specific configuration model
networks with a Poisson degree distribution (ER network), an exponential 
degree distribution, a ternary degree distribution, a Zipf degree distribution
and a power-law degree distribution (scale-free network).
In Sec. \ref{Sec:PercGC} 
we demonstrate the
consistency of our results in the sense that, {\it conditioned} on a 
node belonging to the giant component of a configuration model network,
it cannot belong to a finite tree of any size. 
In Sec. \ref{Sec:ApplGC} we provide examples
illustrating the quality of various approximate descriptions of the degree 
statistics conditioned on the giant component in the 
calculation of the distribution of shortest path lengths,
in the computation of the spectra of sparse 
matrices defined on the giant componen
and in the analysis of epidemic spreading on the giant component.
We conclude the paper with a summary and discussion, in Sec.
\ref{Sec:Summary}.

\section{The Configuration Model Network and its percolation properties}
\label{Sec:CM}

The configuration model is a maximum entropy ensemble of
networks under the condition that the degree distribution
is imposed
\cite{Newman2001,Newman2010}.
Here we focus on the case of undirected networks, in 
which all the edges are bidirectional and
the degree distribution 
$P(k) \equiv P(K=k)$,
$k=0,1,\dots,N-1$,
satisfies
$\sum_k P(k)=1$.
The mean degree over the ensemble of networks is denoted by

\begin{equation}
c=\langle K \rangle=\sum_{k=0}^{N-1} k P(k).
\end{equation}

To construct such a network of a given size, $N$, one can draw
the degrees of all $N$ nodes from 
$P(k)$,
producing the degree sequence
$k_i$, $i=1,\dots,N$
(where $\sum k_i$ must be even).
Before proceeding to the next step of actually constructing the 
network, one should check whether the resulting sequence is
graphic, namely admissible as a degree sequence of at least one
network instance.
The graphicality of the sequence is tested using the Erd{\H o}s-Gallai theorem,
which states that an ordered sequence of the form
$k_1 \ge k_2 \ge \dots, k_N$
is graphic if and only if the condition
\cite{Erdos1960b,Choudum1986}

\begin{equation}
\sum_{i=1}^n k_i \le n(n-1) + \sum_{i=n+1}^N \min (k_i,n)
\end{equation}

\noindent
holds for all the values of $n$ in the range
$1 \le n \le N-1$.

A convenient way to construct a configuration model network 
with a given degree sequence,
$k_i$, $i=1,\dots,N$,
is to prepare the $N$ nodes such that each node, $i$, is 
connected to $k_i$ half edges or stubs
\cite{Newman2010}.
Pairs of half edges 
from different nodes
are then chosen randomly
and are connected to each other in order
to form the network. 
The result is a network with the desired degree sequence and
no correlations.
Note that towards the end of the construction
the process may get stuck.
This may happen in case that the only remaining pairs of stubs
belong to the same node or to nodes which are already connected to each other.
In such cases one may perform some random reconnections 
in order to enable completion of the construction.

A special case of the configuration model is the ER network ensemble,
which is a maximum entropy ensemble under the condition that the
mean degree $c = \langle K \rangle$ is constrained.
ER networks can be constructed by independently connecting
each pair of nodes with probability
$p = c/(N-1)$.
In the thermodynamic limit the resulting degree distribution follows a
Poisson distribution of the form

\begin{equation}
P(k) = \frac{e^{-c} c^k}{k!}.
\label{eq:poisson}
\end{equation}

Consider a random network in the configuration model class, 
described by a degree distribution $P(k)$,  where $k \ge 0$. 
It is well known 
\cite{Newman2001} 
that the fraction $g$ of nodes that reside on the giant component of 
such network can be found using generating functions as described below. 
Let us first introduce the degree generating function
of $P(k)$, namely

\begin{equation}
G_0(x) = \sum_{k=0}^{\infty}  x^k   P(k),
\label{eq:G0}
\end{equation}

\noindent
while
 
\begin{equation}
G_1(x)= \sum_{k=1}^{\infty}  x^{k-1}  \frac{k}{c} P(k) 
\label{eq:G1}
\end{equation}
 
\noindent
is the generating function of the 
distribution of degrees of nodes reached via a random edge.
From the definitions of $G_0(x)$ and $G_1(x)$ in 
Eqs. (\ref{eq:G0}) and (\ref{eq:G1}), respectively,
we find that $G_0(1) = 1$ and $G_1(1) = 1$.

To obtain the probability, $g$, that a random node in the network resides on the giant component, 
one needs to first calculate the probability $\tilde g$ that a random neighbor of a
random node, $i$, belongs to the giant component of the reduced network, which does not include
the node $i$. In the thermodynamic limit, $N \rightarrow \infty$, 
the probability $\tilde g$ is given as a solution of the 
self-consistency equation 
\cite{Molloy1995,Molloy1998}

\begin{equation}
1 - \tilde g = G_1(1-\tilde g).
\label{eq:tg-fpe}
\end{equation}

\noindent
The left hand side of this equation is the probability that a random neighbor of a random
reference node in the network
does not reside on the giant component of the reduced network from which the
reference node is removed. 
The right hand side represents the same quantity
in terms of its neighbors, namely as the probability that none of the neighbors of such node
resides on the giant component of the reduced network. 
Once $\tilde g$ is known, the probability $g$ can be obtained from

\begin{equation}
g = 1 -  G_0(1 - \tilde g),
\label{eq:g}
\end{equation}

\noindent
This relation is based on the same consideration as Eq. (\ref{eq:tg-fpe}), where the difference is that the
reference node is a random node rather than a random neighbor of a random node.

Clearly, $\tilde g=0$ 
is always a solution 
of Eq. (\ref{eq:tg-fpe}). 
A random network exhibits a giant component, 
if Eq. (\ref{eq:tg-fpe}) 
also has a non-trivial solution.
The condition for the existence of a giant component can 
be expressed in the form
\cite{Newman2010}

\begin{equation}
\sum_{k=2}^{\infty} \frac{k (k-1)}{c} P(k)
= \frac{\langle K^2\rangle -\langle K\rangle }{\langle K \rangle} > 1,
\end{equation}

\begin{figure}
\begin{center}
\includegraphics[width=0.45\textwidth]{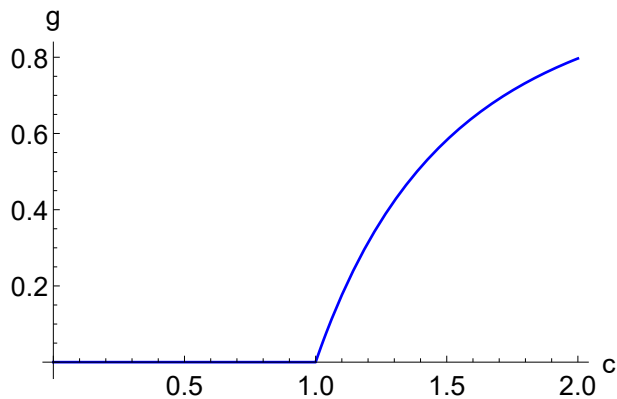} 
\end{center}
\caption{
(Color online)
The probability $g$ that a random node
in an ER network
resides on the 
giant component,
as a function of the mean degree, $c$.
For $c<1$ there is no giant component and thus $g=0$.
At $c=1$ there is a percolation transition, above which
$g$ increases monotonically towards the dense network
limit of $g=1$.
}
\label{fig:1}
\end{figure}

\noindent
which is known as the Molloy-Reed criterion \cite{Molloy1995,Molloy1998}. 
In essence, this criterion states that a giant component exists 
if the mean excess degree of the neighbours of a random node exceeds one.
In Fig. \ref{fig:1} we present the parameter $g$, which is the probability that 
a random node resides
on the giant component, for an ER network, as a function of the mean degree, $c$.
For $c<1$ there is no giant component and thus $g=0$.
The percolation transition takes place at $c=1$, above which
$g$ gradually increases towards the dense network limit
of $g=1$.

\section{The Degree Distribution on the Giant Component}
\label{Sec:GC}

The probability, $g$, that a randomly selected node belongs to the giant component
can be expressed in the form

\begin{equation}
g = \sum_{k=0}^{\infty}  g_k  P(k),
\end{equation}

\noindent
where $g_k$ is
the {\it conditional probability} that a random node belongs to 
the giant component, {\it given} that its degree is $k$. 
Comparing this expression to Eq. (\ref{eq:g}) we find that

\begin{equation}
g_k = 1 - (1-\tilde g)^k .
\end{equation}

\noindent
To make the conditioning explicit, we introduce an indicator variable 
$\Lambda \in \{0,1\}$, with $\Lambda=1$ indicating that an event happens on 
the giant component, whereas $\Lambda=0$ indicates that it happens on one 
of the finite components of the network. 
The probability that a random node resides on the giant component is
given by $P(\Lambda=1)=g$, while the probability that it resides on one of
the finite components is $P(\Lambda=0)=1-g$.
The probability
that a random node of a given degree $k$ resides on the giant component 
is given by

\begin{equation}
P(\Lambda=1|K=k) =  g_k  = 1 - (1-\tilde g)^k,
\end{equation}

\noindent
while the probability that it resided on one of the finite components is

\begin{equation}
P(\Lambda=0|K=k) =  1 - g_k  = (1-\tilde g)^k.
\end{equation}

\noindent
Using Bayes' theorem 

\begin{equation}
P(K=k|\Lambda=\lambda) 
= 
\frac{P(\Lambda=\lambda | K=k)}{P(\Lambda=\lambda)}  P(K=k),
\end{equation}

\noindent
we invert these relations so as to obtain the degree distributions 
conditioned on a node to belong to the giant and the finite 
components, respectively.
For brevity, in the rest of the paper we use a more compact notation,
in which $P(K=k)$, 
$P(\Lambda=\lambda)$ 
and 
$P(K=k | \Lambda=\lambda)$
are replaced by
$P(k)$,
$P(\lambda)$
and
$P(k|\lambda)$,
respectively,
except for a few places in which the more detailed notation is needed for clarity.
The conditional degree distribution of nodes which reside on the giant component
is given by

\begin{equation}
P(k|1) = \frac{1-(1-\tilde g)^k}{g} P(k),
\label{eq:pk1}
\end{equation}

\noindent
while the conditional degree distribution for nodes which reside
on the finite tree components is

\begin{equation}
P(k|0) = \frac{(1-\tilde g)^k}{1-g} P(k).
\label{eq:pk0}
\end{equation}

\noindent
This result can be expressed in the form

\begin{equation}
P(k|0) = \frac{1}{1-g} e^{- a k} P(k),
\label{eq:pk0e}
\end{equation}

\noindent
where 
$a= - \ln(1 - \tilde g)$.
This highlights the fact that the degree distribution of the
finite components is an exponentially attenuated variant of the original degree
distribution.
The result for the finite components, 
Eq. (\ref{eq:pk0}), 
was first derived by Molloy and Reed 
\cite{Molloy1995,Molloy1998} (although in a less transparent form), 
while the result for the giant 
component, 
Eq. (\ref{eq:pk1}), 
was reported by Engel et al. 
\cite{Engel2004} 
for the special case of ER networks 
(for which $\tilde g = g$).

The mean degree conditioned on the giant component is

\begin{equation}
c_1 = \mathbb{E}[ K | 1 ] = \sum_{k=1}^{\infty} k P(k|1).
\label{eq:c1}
\end{equation}

\noindent
Using Eq. (\ref{eq:pk1}) 
and performing the summation, we obtain

\begin{equation}
c_1 = \frac{c}{g} \left[ 1-(1-\tilde g)^2 \right] = \frac{  (2-\tilde g) \tilde g }{g} c.
\label{eq:c_1}
\end{equation}

\noindent
The mean degree conditioned on the finite components
is given by

\begin{equation}
c_0= \mathbb{E}[ K |  0 ] = \sum_{k=0}^{\infty} k P(k|0).
\label{eq:c_0p}
\end{equation}

\noindent
From Eq. (\ref{eq:pk0}),
we obtain

\begin{equation}
c_0
= \frac{ (1-\tilde g)^2 }{1-g} c.
\label{eq:c_0}
\end{equation}

\noindent
Using Eq. (\ref{eq:pk0}) and the generating function $G_1(x)$,
it can be shown that $(1-\tilde g)^2/(1-g) < 1$ for
$0 < \tilde g < 1$, which implies that
$c_0 < c$ and $c_1 > c$.
For ER networks these results specialize to 
$c_1= (2-g)c$ 
and 
$c_0 = (1-g)c$, 
respectively. 
Actually, the value of $c_0$ corresponds to the mean degree
below the percolation threshold 
as it must correspond to a sub-percolating configuration model.

In the literature, the finite component result of
Eq. (\ref{eq:pk0}) 
is referred to as a discrete duality relation 
\cite{Molloy1998,Bollobas2007,Janson2010}. 
Indeed for ER networks $P(k|0)$ is in itself a Poisson distribution
of the form

\begin{equation}
P(k|0) = \frac{ e^{-c_0} c_0^k }{k!},
\end{equation}

\noindent
where 
$c_0= c(1-g)$ 
is the mean degree of the nodes which reside on the finite components. 
The degree distribution, 
restricted to nodes on the finite components of an ER network is thus 
of the same type as the degree distribution of the entire network, 
albeit with a renormalized parameter for the mean degree, $c_0$. 
Note that $c_0<1$ for any $c>1$, reflecting the fact that the finite 
components are equivalent to a sub-percolating ER network.

An analogous {\it parametric} renormalization relating the degree 
distribution of the whole network to a degree distribution conditioned on the finite components 
is found for any degree distribution which has a component which scales
exponentially in $k$.
Such degree distributions can be expressed in the form

\begin{equation}
P(k) = \phi(k) e^{- \alpha k},
\label{eq:exp}
\end{equation}

\noindent
where $\alpha > 0$, and the function $\phi(k) \ge 0$ is chosen such that 
$P(k)$ is properly normalized. 
Clearly, the simplest example of such degree distribution 
is the exponential distribution, for which 
$\phi(k) = 1 - e^{-\alpha}$
is merely a normalization constant. 
For networks with an exponential component in the degree distribution 
as described 
in Eq. (\ref{eq:exp}), 
the degree distribution conditioned on the finite components
takes the form

\begin{equation}
P(k|0) = \frac{1}{1-g} \phi(k) e^{- \alpha_0 k}
\end{equation}

\noindent
with 
$\alpha_0 = \alpha - \ln (1-\tilde g)$.
This simple parametric renormalization with respect to 
Eq. (\ref{eq:exp}) 
is in close analogy to the results obtained earlier for ER networks. 
The degree distribution conditioned on the giant component can 
be compactly expressed as

\begin{equation}
P(k|1) = \frac{ e^{- \alpha k} - e^{- \alpha_0 k} }{g} \phi(k).
\end{equation}

In order to obtain the conditional degree distributions for a
given network, one needs to evaluate the parameters
$g$ and $\tilde g$. 
The latter is obtained from the solution of 
Eq. (\ref{eq:tg-fpe}), 
while the former is obtained by inserting the solution for $\tilde g$ 
into (\ref{eq:g}).

\section{Degree-Degree Correlations on the Giant Component}
\label{Sec:JointGC}

Having computed degree distributions conditioned on the giant and 
finite components of configuration model networks, we now turn to investigating 
the micro-structure of these giant and finite components further by 
looking at various joint degree distributions and degree-degree correlation. 
We shall find that --- on the giant component --- there are degree-degree 
correlations of any order. This could have been anticipated, as 
degree-degree correlations of arbitrarily high order are clearly 
required in order to exclude the possibility that a randomly selected node 
belongs to a tree of {\it any finite size}. 
In what follows we go some way to quantify these correlations.
The key step is to use 
Eq. (\ref{eq:tg-fpe}) 
to express the powers 
$(1-\tilde g)^k$ 
appearing in 
Eq. (\ref{eq:g}), 
resulting in

\begin{equation}
g = \sum_{ k;\{k_{\mu} \} }  
\left[1-\prod_{\mu=1}^k (1- \tilde g)^{k_\mu-1}\right]
P(k) 
\prod_{\mu=1}^k \frac{k_\mu}{c} P(k_\mu).
\label{eq:gkkmu}
\end{equation}

\noindent
Here we use the notation
$k;\{ k_{\mu} \}$
to denote a configuration consisting of 
a central node of degree $k$, surrounded by a first coordination 
shell of nodes with degrees 
$k_\mu, \mu=1,\dots,k$.
The probabilistic interpretation of this identity is that the probability
that a random node of degree $k$, whose neighbors are of degrees
$k_1,\dots,k_k$ resides on the giant component is

\begin{equation}
P(1|k; \{ k_{\mu} \}) 
= 
\left[1-\prod_{\mu=1}^k (1- \tilde g)^{k_\mu-1}\right],
\end{equation}

\noindent
while the probability that it resides on one of the finite components is

\begin{equation}
P(0|k; \{ k_{\mu} \})= \prod_{\mu=1}^k (1- \tilde g)^{k_\mu-1}.
\end{equation}

\noindent
Using Bayes' Theorem, one can 
invert these relations to obtain

\begin{equation}
P(k; \{ k_{\mu} \}|1) 
= 
\frac{1}{g} \left[1-\prod_{\mu=1}^k (1- \tilde g)^{k_\mu-1}\right]
P(k) \prod_{\mu=1}^k \frac{k_\mu}{c} P( k_{\mu} ),
\label{eq:pk;kmu1}
\end{equation}

\noindent
and

\begin{equation}
P(k; \{ k_{\mu} \}|0) 
= 
\frac{1}{1-g}
\left[\prod_{\mu=1}^k (1- \tilde g)^{k_\mu-1}\right] 
P(k)  \prod_{\mu=1}^k \frac{k_\mu}{c} P( k_{\mu}),
\label{eq:pk;kmu0}
\end{equation}

\noindent
as the probabilities for nodes to have a degree $k$ and first 
neighbour shell configuration 
$k_1,\dots,k_k$, 
conditioned 
on this happening on the giant component, and on one of the finite components,
respectively. 
Note that 
Eq. (\ref{eq:pk1}) 
correctly 
predicts that the probability of a node of degree $k=0$ to 
belong to the giant component is zero. 
Moreover, 
Eq. (\ref{eq:pk;kmu1}) 
also correctly 
predicts that the probability of a node of degree $k$ to connect to 
$k$ nodes of degree $1$, thereby forming an isolated $(k+1)$-star, is 
zero on the giant component.

Marginalizing $P(k; \{ k_{\mu} \}|1)$, namely summing 
Eq. (\ref{eq:pk;kmu1}) 
over all the values of 
$k_2,\dots,k_k$,
and replacing $k_1 \rightarrow k'$,
gives 
the probability of a random
node of degree $k$ to be connected to a node of degree $k'$, 
{\it conditioned} on them being on the giant component 

\begin{equation}
P(k;k'|1)= \frac{1}{g} \left[ 1-(1-\tilde g)^{k-1}(1-\tilde g)^{k'-1} \right]
P(k) \frac{k'}{c} P(k'),  \quad k \ge 1.
\label{eq:pkkpg1}
\end{equation}

\noindent
Similarly, from
Eq. (\ref{eq:pk;kmu0})
we obtain
the probability of a random
node of degree $k$ to be connected to a node of degree $k'$, 
under the condition that they do not reside on the giant component 

\begin{equation}
P(k;k'|0)= \frac{1}{1-g}
(1-\tilde g)^{k-1} (1-\tilde g)^{k'-1} 
P(k)  \frac{k'}{c} P(k'),
\quad k \ge 1.
\label{eq:pkkpg0}
\end{equation}

\noindent
Here we have exploited the 
fact that the $k_2,\dots,k_k$ averages 
over the distributions of
neighbouring degrees factor, 
each of them giving

\begin{equation}
\sum_{k_\mu}   (1-\tilde g)^{k_\mu-1}     \frac{k_\mu}{c} P(k_{\mu}) 
= 1-\tilde g,
\end{equation}

\noindent
by using
Eq. (\ref{eq:tg-fpe}).
Marginalizing Eqs. (\ref{eq:pkkpg1}) and (\ref{eq:pkkpg0}) by summing 
over $k'$, we recover $P(k|1)$ and $P(k|0)$ as given by 
Eqs. (\ref{eq:pk1}) and (\ref{eq:pk0}). 
On the other hand, marginalizing 
Eq. (\ref{eq:pkkpg1}) by summing over $k \ge 1$, gives the probability, 
starting from a randomly chosen node on the giant component, 
to reach a node of degree $k'$, namely

\begin{equation}
\widetilde P(k'|1) 
\equiv 
\sum_{k \ge 1} P(k;k'|1).
\end{equation}

\noindent
Carrying out the summation we obtain

\begin{equation}
\widetilde P(k'|1) 
= \frac{1}{g}
\left[1-p_0 - \frac{1-g-p_0}{1-\tilde g}
(1-\tilde g)^{k'-1}\right] \frac{k'}{c} P(k'),
\label{eq:p1k}
\end{equation}

\noindent
where $p_0 = P(K=0)$, namely the probability of an isolated node 
in the original network.
It is easy to see that this is a normalized distribution.
It is also important to stress the asymmetric role of the two 
degrees $k$ and $k'$ appearing in 
Eqs. (\ref{eq:pkkpg1}) 
and (\ref{eq:pkkpg0}). 

Consider a random edge in a configuration model network.
The joint degree distribution,
$\widehat P(k,k')$,
of the nodes which reside on both sides
of such edge is given by

\begin{equation}
\widehat P(k,k') = \frac{k}{c} P(k) \frac{k'}{c} P(k').
\end{equation}

\noindent
The non-giant components of a configuration model network
constitute a sub-network which is itself a configuration model network, 
and is in the sub-percolation regime.
The degree distribution, $P(k|0)$, of this sub-network
is given by 
Eq. (\ref{eq:pk0}).
Thus, the joint degree distribution 
of pairs of connected nodes which reside
on the non-giant components
is given by

\begin{equation}
\widehat P(k,k' |0) =
\frac{(1-\tilde g)^{k-1} (1-\tilde g)^{k'-1}}{(1-\tilde g)^2} 
\frac{k}{c} P(k) \frac{k'}{c} P(k').
\label{eq:Pkk'0}
\end{equation}

\noindent
The fraction of edges in the network which reside on the giant component is
denoted by $g_E$.
It is given by

\begin{equation}
g_E = 1 - (1-\tilde g)^2,
\end{equation}

\noindent
while the fraction of edges which reside on the non-giant components is 
$1 - g_E = (1 - \tilde g)^2$.
Therefore, the joint degree distribution can be expressed in the form

\begin{equation}
\widehat P(k,k') = 
\left[ 1 - (1 - \tilde g)^2 \right] \widehat P(k,k'|1)
+
(1 - \tilde g)^2 \widehat P(k,k'|0).
\label{eq:Pkk'}
\end{equation}

\noindent
Using Eqs. (\ref{eq:Pkk'0}) and (\ref{eq:Pkk'})
we find that

\begin{equation}
\widehat P(k,k'|1) = 
\frac{1 - (1-\tilde g)^{k+k'-2} }{1 - (1-\tilde g)^2} 
\frac{k}{c} P(k) \frac{k'}{c} P(k').
\label{eq:Pkk1'}
\end{equation}

\section{Assortativity on the Giant Component} 
\label{Sec:assort}

From the joint probability 
$\widehat P(k,k'|1)$ 
that a randomly chosen edge on the 
giant component connects two vertices of degrees $k$ 
and $k'$, one obtains the corresponding probability 
for a random edge to connect nodes of 
{\it excess-degrees} 
$k$ and $k'$,
by a simple shift of arguments as

\begin{equation}
\widehat P_e(k,k'|1) = \widehat P(k+1,k'+1|1).
\end{equation}

\noindent
The conditional joint probability of excess degrees, 
$\widehat P_e(k,k'|1)$,
is given by

\begin{equation}
\widehat P_e(k,k'|1) 
= 
\left[ \frac{1 - (1-\tilde g)^{k+k'-2} }{1 - (1-\tilde g)^2} \right]
\frac{k+1}{c} P(k+1)
\frac{k'+1}{c} P(k'+1).
\label{eq:Pkk'1}
\end{equation}

\noindent
Summing over $k$, we obtain the marginal distribution

\begin{equation}
\widehat P_e(k|1) = 
\left[ \frac{1 - (1-\tilde g)^{k-1} }{1 - (1-\tilde g)^2} \right]
\frac{k+1}{c} P(k+1).
\end{equation}

\noindent
In terms of these definitions, the assortativity 
coefficient on the giant component is given by 
\cite{Newman2002b}

\begin{equation}
r
=
\frac{1}{\hat \sigma^2} 
\sum_{k,k'\ge 0} k k'
\left[\widehat P_e(k,k'|1) 
- \widehat P_e(k|1) \widehat P_e(k'|1)\right],
\label{eq:r}
\end{equation}

\noindent
where 

\begin{equation}
\hat \sigma^2 = 
\sum_{k \ge 0} k^2 \widehat P_e(k|1) -  \left[ \sum_{k \ge 0} k \widehat P_e(k|1) \right]^2
\end{equation}

\noindent
is the
variance of 
$\widehat P_e(k|1)$. 
The assortativity coefficient is actually the
Pearson correlation coefficient of degrees between pairs
of linked nodes
\cite{Newman2002b}.

While the assortativity coefficient can be evaluated 
directly from Eq. (\ref{eq:r}), it turns out that
there is a more effective approach for its calculation,
using generating functions. 
To this end, we introduce the bivariate
generating function, 
$B(u,v)$,
of 
$\widehat P_e(k,k'|1)$,
which is given by

\begin{equation}
B(u,v) 
= 
\sum_{k,k' \ge 0} \widehat P_e(k,k'|1) u^k v^{k'}.
\label{eq:E}
\end{equation}

\noindent
This function is symmetric in $u$ and $v$,
reflecting the symmetric form of 
$\widehat P_e(k,k'|1)$
in terms of $k$ and $k'$.
We also introduce the generating function,
$S(u)$, of the marginal distribution
$\widehat P_e(k|1)$,
which takes the form

\begin{equation}
S(u) = \sum_{k \ge 0} \widehat P(k|1) u^k. 
\end{equation}

\noindent
Note that it can be expressed in terms of the
bivariate generating function, as $S(u) = B(u,1)$.
Inserting the expression for $\widehat P_e(k,k'|1)$ from
Eq. (\ref{eq:Pkk'1}) into 
Eq. (\ref{eq:E}),
we find that the bivariate generating function $B(u,v)$
can be expressed in terms of the generating function
$G_1(x)$ of the degree distribution $P(k)$.
It takes the form

\begin{equation}
B(u,v) =
\frac{G_1(u) G_1(v) - G_1[(1-\tilde g) u] G_1[(1-\tilde g) v]}{1 - (1-\tilde g)^2}.
\label{eq:E2}
\end{equation}

\noindent
Plugging in $v=1$ we obtain

\begin{equation}
S(u) = \frac{G_1(u) - (1-\tilde g) G_1[(1-\tilde g) u]}{1 - (1-\tilde g)^2}.
\label{eq:S}
\end{equation}

\noindent
Expressing the terms on the right hand side of Eq. (\ref{eq:r})
in terms of derivatives of the generating functions $B(u,v)$ and $S(u)$,
we express the assortativity coefficient 
in the form

\begin{equation}
r 
= 
\frac{\partial_u \partial_v B(u,v) -[\partial_u S(u)]^2}
{(u \partial_u)^2 S(u) - [\partial_u S(u)]^2}\,\Bigg|_{u=v=1}.
\label{eq:R}
\end{equation}

\noindent
In the next section we use this formulation to obtain exact
analytical results for the assortativity coefficients on the
giant components of different configuration model networks.
The assortativity coefficient is expected to be negative, which
implies that the giant component of a configuration model network
is disassortative. 
This is due to the fact that high-degree nodes are over-represented in
the giant component. Thus, in order that the giant component
will be a single connected component, low degree nodes must
have a greater than normal probability to connect to high degree
nodes. In particular, a node of degree $k=1$ on the giant
component must be connected to a node of degree $k' \ge 2$,
while a node of degree $k=2$ can have at most one neighbor of 
degree $k'=1$. The disassortativity of the giant component is
most pronounced just above the percolation threshold, where
most nodes are of low degrees.

\section{Analysis of specific network models}
\label{Sec:examp}

In this section we discuss in detail some specific network models
and the properties of their giant components.

\subsection{Erd{\H o}s-R\'enyi networks}

Consider an ER network of $N$ nodes and mean degree $c>1$.
In this case the degree distribution follows a Poisson
distribution,
given by Eq. (\ref{eq:poisson}).
The generating functions $G_0$ and $G_1$ 
of this distribution coincide
and satisfy 

\begin{equation}
G_0(x) = G_1(x) = e^{-c(1-x)}.
\end{equation}

\noindent
As a result, in this case $\tilde g = g$.
Using Eq. (\ref{eq:g}) one obtains a closed form
expression for $g$, which is given by

\begin{equation}
g = 1 + \frac{W(-ce^{-c})}{c},
\label{eq:gER}
\end{equation}

\noindent
where $W(x)$ is the Lambert W function
\cite{Olver2010}.
In this case a giant component exists for $c>1$ 
(Fig. \ref{fig:1}).
The degree distribution on the giant component 
is obtained from Eq. (\ref{eq:pk1}),
where $g$ and $\tilde g$ are given by Eq. (\ref{eq:gER}).
It is given by

\begin{equation}
P(k |1) = 
\frac{1}{g} \left[ \frac{e^{-c} c^k}{k!}
- (1-g) \frac{ e^{- c(1-g)} [c(1-g)]^k}{k!} \right],
\label{eq:PkER}
\end{equation}

\begin{figure}
\begin{center}
\includegraphics[width=0.3\textwidth]{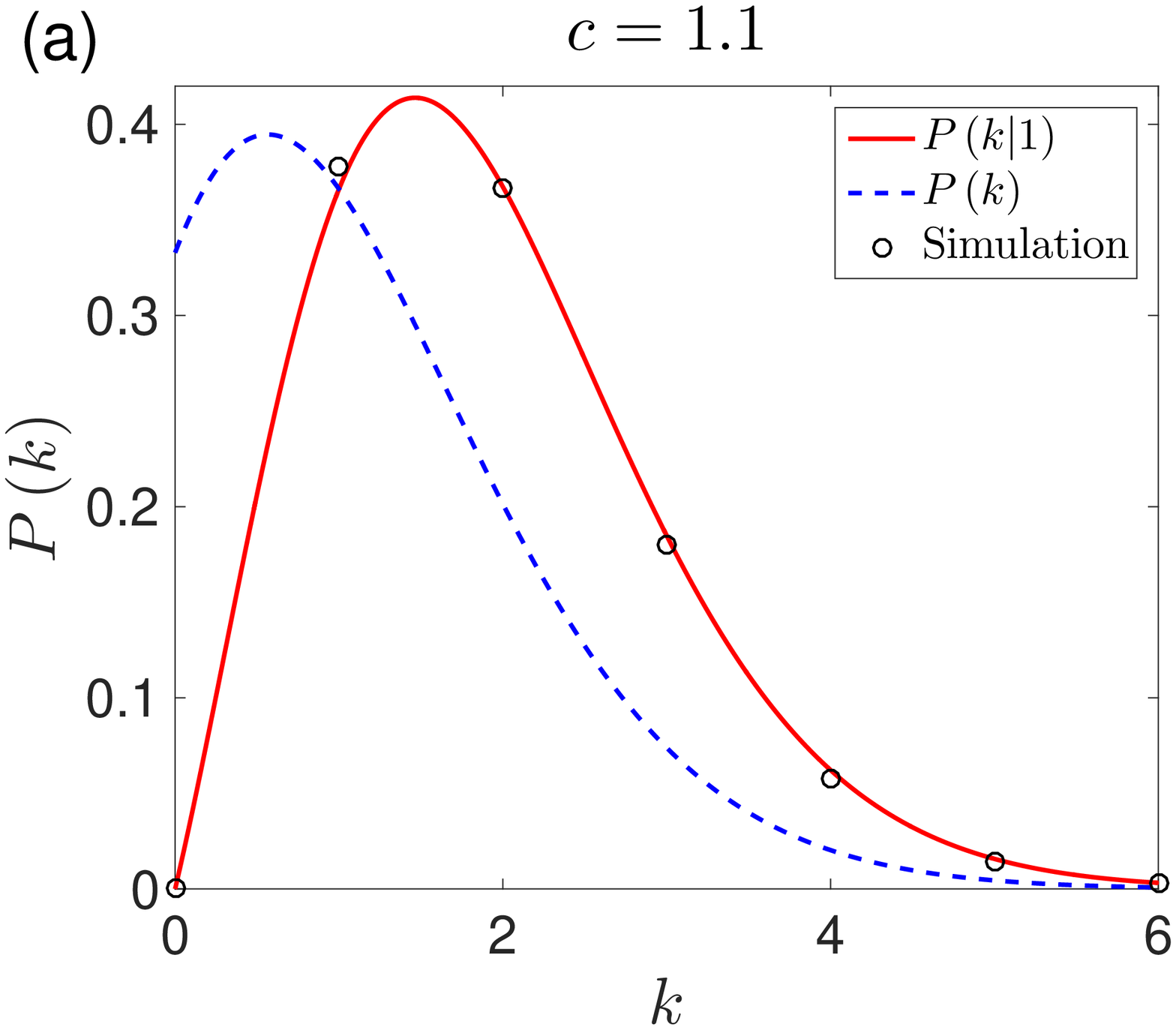} \hfil
\includegraphics[width=0.3\textwidth]{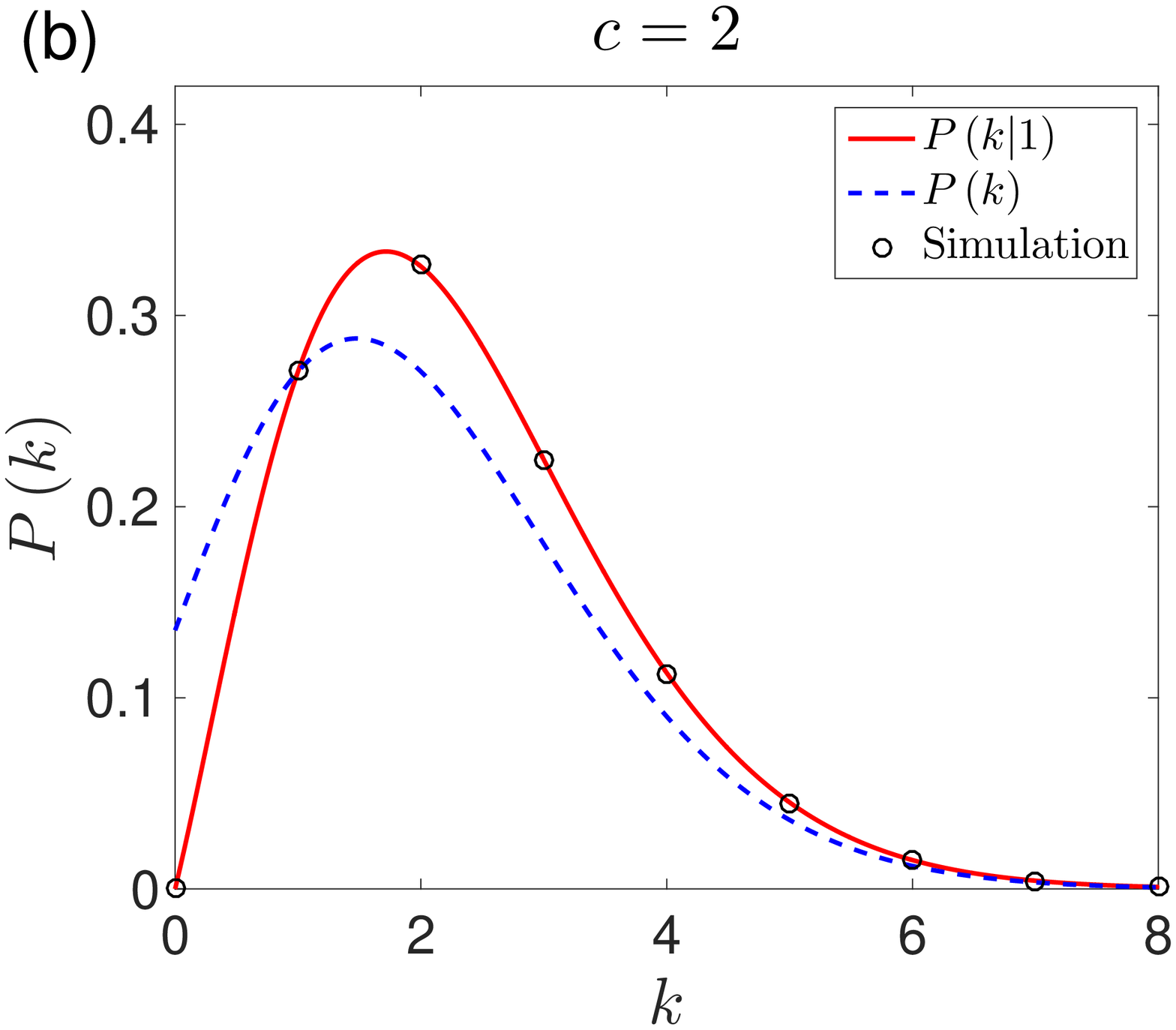} \hfil
\includegraphics[width=0.3\textwidth]{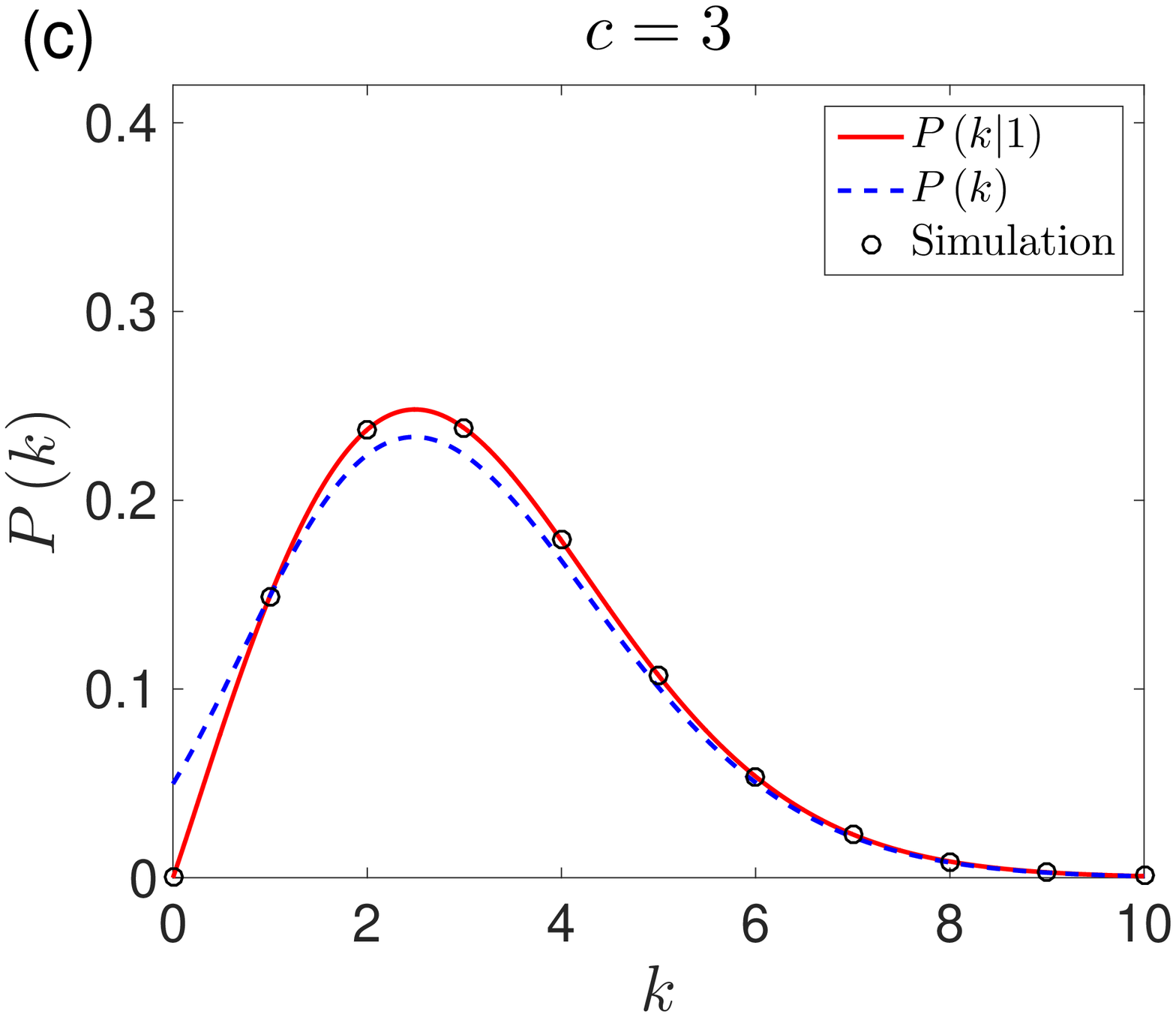} 
\end{center}
\caption{
(Color online)
Analytical results for the degree distribution,
$P(k|1)$, 
of the giant 
component of an ER network (solid lines),
obtained from Eq. (\ref{eq:PkER}),
for 
$c=1.1$ (a), 
$c=2$ (b) 
and 
$c=3$ (c).
The analytical results are in excellent agreement with the
results of computer simulations (circles), for a network
of $N=1000$ nodes.
For comparison, the degree distribution, $P(k)$, of the 
whole network,
obtained from 
Eq. (\ref{eq:poisson}) 
is also shown (dashed lines).
}
\label{fig:2}
\end{figure}

\noindent
which takes the form of the difference between two Poisson distributions.
In Fig. \ref{fig:2} 
we present analytical results for the degree distribution,
$P(k|1)$, of the giant components of ER networks with mean degrees
$c=1.1$, $2$ and $3$ (solid lines),
obtained from Eq. (\ref{eq:PkER}).
The analytical results are found to be in excellent agreement with
the results of computer simulations (circles),
for a network of size $N=1000$.
For comparison we also show the degree distribution
on the entire network (dashed lines),
obtained from Eq. (\ref{eq:poisson}).

The mean degree of the giant component of an ER network,
obtained from Eq. (\ref{eq:c1}),
is given by

\begin{equation}
\mathbb{E}[K|1] =
(2-g) c,
\label{eq:E[K|1]ER}
\end{equation}

\noindent
while the mean degree of the finite components,
obtained from Eq. (\ref{eq:c_0}) is

\begin{equation}
\mathbb{E}[K|0] =
(1-g) c.
\label{eq:E[K|0]ER}
\end{equation}

\begin{figure}
\begin{center}
\includegraphics[width=0.45\textwidth]{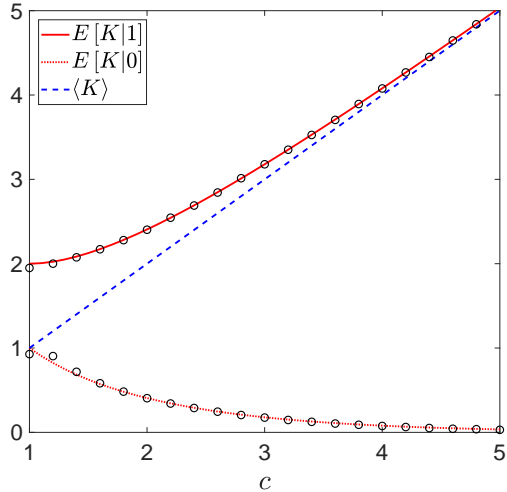} 
\end{center}
\caption{
(Color online)
Analytical results for the mean degree of the giant component, 
$\mathbb{E}[K|1]$, 
of an ER network (solid line),
as a function of $c$, for $c \ge 1$,
obtained from 
Eq. (\ref{eq:E[K|1]ER}). 
The analytical results are found to be in excellent agreement with the
results of computer simulations (circles),
for a network of $N=1000$ nodes.
For comparison, the mean degree, 
$\langle K \rangle =c$, 
of the whole network is also shown
(dashed line).
The mean degree of the giant component starts from 
$\mathbb{E}[K|1]=2$ at $c=1$ 
and gradually approaches the result for the 
whole network as $c$ is
increased. In contrast, the mean 
degree, 
$\mathbb{E}[K|0]$,
of the finite components
(dotted line)
starts from $\mathbb{E}[K|0]=1$ at $c=1$
and gradually decreases to zero as $c$ is increased.
}
\label{fig:3}
\end{figure}

\noindent
In Fig. \ref{fig:3}
we present analytical results for the mean degree, 
$\mathbb{E}[K|1]$, 
of the giant component 
(solid line)
and the mean degree, 
$\mathbb{E}[K|0]$, 
of the finite components 
(dotted line),
of an ER network
as a function of $c$. 
The mean degree of the whole network, 
$\langle K \rangle=c$, 
is also shown (dashed line).
It is observed that at the percolation threshold ($c=1$), 
$\mathbb{E}[K|1]=2$, while
$\langle K \rangle =1$.
As $c$ is increased, the mean degree of the 
giant component converges asymptotically towards to 
overall mean degree of the network,
while the mean degree of the finite components decays to zero.

The assortativity coefficient for the giant component of an 
ER network is given by

\begin{equation}
r = - \frac{c(1-g)^2}{(2-g)^3 - (1-g)(2-g) - c(1-g)^2}.
\label{eq:rER}
\end{equation}

\noindent
In the limit of large $c$, the index $r$ decreases according to
$r \simeq -c e^{-2c}$.
For values of $c$ just above the percolation threshold,
we find that 

\begin{equation}
r \simeq - \frac{1}{5} + \frac{12}{25} (c-1)^2 
+ {\mathcal O} \left[ (c-1)^3 \right].
\end{equation}

\begin{figure}
\begin{center}
\includegraphics[width=0.45\textwidth]{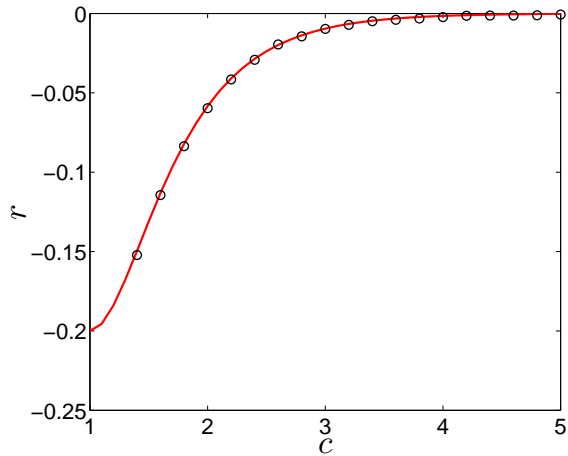} 
\end{center}
\caption{
(Color online)
The assortativity coefficient $r$ of the giant component of an ER network,
as a function of $c$.
The analytical results (solid line), obtained from 
Eq. (\ref{eq:rER}),
are found to be in excellent agreement with the results of computer 
simulations (circles), performed for networks of size $N=5000$.
The negative values of $r$ imply that the giant component is disassortative,
namely high degree nodes preferentially connect to low degree nodes
and vice versa.
As $c$ is increased, the giant component encompasses an increasing fraction
of the entire network, and $r$ decays to zero.
}
\label{fig:4}
\end{figure}

\noindent
The negative value of $r$ implies that the giant component 
is disassortative, meaning that high degree 
nodes on the giant component tend to 
connect to low degree nodes and vice versa.
As $c$ is increased, the absolute value of $r$ gradually decreases.
In Fig. \ref{fig:4} we present the assortativity coefficient, $r$, 
of the giant component of an ER network, as a function of $c$.
Just above the percolation transition, the assortativity coefficient is large and
negative. Its absolute value gradually decreases and eventually vanishes 
as $c$ is increased, reflecting the fact that the giant component 
coincides with the entire network, and all the correlations are lost.

\subsection{Configuration model networks with an exponential degree distribution}

Consider a configuration model network with an exponential degree
distribution of the form

\begin{equation}
P(k) = A e^{- \alpha k},
\end{equation}

\noindent
where $k \ge k_{\rm min}$. 
Here we focus on the case of 
$k_{\rm min}=1$, 
for which the normalization 
factor is $A=e^{\alpha} - 1$.
The mean degree is given by

\begin{equation}
c=\langle K \rangle = \frac{1}{1 - e^{- \alpha}}.
\end{equation}

\noindent
For the analysis presented below, it is convenient to parametrize the
degree distribution in terms of the mean degree, $c$.
Plugging in 
$\alpha = \ln c - \ln (c-1)$ 
we obtain

\begin{equation}
P(k) = \frac{1}{c} \left( \frac{c-1}{c} \right)^{k-1},
\end{equation}

\noindent
with $k \ge 1$. The degree generating function $G_0(x)$ is given by

\begin{equation}
G_0(x)
= \frac{x}{c-x(c-1)}.
\end{equation}

\noindent
It exhibits two trivial fixed points, namely
$G_0(0)=0$ and $G_0(1)=1$.
The cavity generating function $G_1(x)$ is

\begin{equation}
G_1(x) 
= \frac{1}{\left[ c - x(c-1) \right]^2}.
\end{equation}

\noindent
This generating function has a trivial fixed point given by $G_1(1)=1$.
The size of the giant component is obtained using a two step process.
In the first step we find the non-trivial fixed point
of $G_1(x)$, by solving Eq. (\ref{eq:tg-fpe}) for $\tilde g$.
We find that

\begin{equation}
\tilde g = \frac{c-3}{2(c-1)} + \frac{1}{2} \sqrt{ \frac{c+3}{c-1} }.
\label{eq:tildegexp}
\end{equation}

\noindent
In the second step we obtain the fraction of nodes which reside
on the giant component, which is given by Eq. (\ref{eq:g}), namely

\begin{equation}
g =  \frac{3c}{2(c-1)} 
- \frac{c (c+3)^{1/2} }{2 (c-1)^{3/2} }.
\label{eq:gexp}
\end{equation}

\noindent
The percolation transition occurs at $c=3/2$, such that a giant component
exists for $c>3/2$. 
The degree distribution on the giant component,
obtained from 
Eq. (\ref{eq:pk1}),
takes the form

\begin{equation}
P(k|1) = \left[ \frac{   1 - (1-\tilde g)^k     }{(c-1)g} \right] 
\left( \frac{c-1}{c} \right)^k,
\end{equation}

\noindent
where $\tilde g$ is given by
Eq. (\ref{eq:tildegexp})
and $g$ is given by
Eq. (\ref{eq:gexp}).
The mean degree on the giant component is
given by Eq. (\ref{eq:c_1}).
The assortativity coefficient of a configuration model with an exponential
degree distribution takes the form

\begin{equation}
r =  - 
\frac{2(c-1)(1-\tilde g)^2 [1-(1-\tilde g)^{3/2}]^2}
{[1-(1-\tilde g)^2] \{ 1 - (1-\tilde g)^{7/2} + 3 (c-1)[1-(1-\tilde g)^5] \}
- 2 (c-1)[1-(1-\tilde g)^{7/2} ]^2 }.
\end{equation}

\noindent
In the limit of large $c$, the assortativity 
coefficient decreases to zero according to 

\begin{equation}
r \simeq -2 c^{-4} + 2c^{-5} + {\mathcal O}(c^{-6}).
\end{equation}

\noindent
Just above the percolation threshold, which is located at $c=3/2$, 
the assortativity coefficient can be approximated by

\begin{equation}
r \simeq -\frac{3}{13} + \frac{2240}{1521} \left( c - \frac{3}{2} \right)^2  
- \frac{35840}{13689} \left( c - \frac{3}{2} \right)^3 
+ {\mathcal O} \left( c - \frac{3}{2} \right)^4.
\end{equation}

\subsection{Configuration model networks with a ternary degree distribution}

The properties of the giant components of random networks are very
sensitive to the abundance of nodes of low degrees, particularly
nodes of degree $k=1$ (leaf nodes) and $k=2$.
Nodes of degree $k=0$ (isolated nodes) are excluded from the
giant component and their weight in the overall degree distribution
has no effect on the properties of the giant component.
Therefore, it is useful to consider a simple configuration model
in which all nodes are restricted to a small number
of low degrees. 
Here we consider a configuration model network with a ternary 
degree distribution of the form
\cite{Newman2010}

\begin{equation}
P(k) = p_1 \delta_{k,1} + p_2 \delta_{k,2} + p_3 \delta_{k,3},
\label{eq:trinary}
\end{equation}

\noindent
where $\delta_{k,n}$ is the Kronecker delta,
and 
$p_1+p_2+p_3=1$.
The mean degree of such network is given by

\begin{equation}
\langle K \rangle = p_1 + 2 p_2 + 3 p_3.
\end{equation}

\noindent
The generating functions are

\begin{equation}
G_0(x) = p_1 x + p_2 x^2 + p_3 x^3,
\label{eq:G0tri}
\end{equation}

\noindent
and

\begin{equation}
G_1(x) = \frac{ p_1  + 2 p_2 x + 3 p_3 x^2}{p_1 + 2 p_2 + 3 p_3}.
\label{eq:G1tri}
\end{equation}

\noindent
Solving Eq. (\ref{eq:tg-fpe}) for $\tilde g$, with $G_1(x)$ given by 
Eq. (\ref{eq:G1tri}), 
we find that

\begin{equation}
\tilde g = 
\begin{dcases}
0 &  \ \ \ \  p_3 \le \frac{p_1}{3} \\
1 - \frac{p_1}{3p_3} &  \ \ \ \  p_3 > \frac{p_1}{3}.
\end{dcases}
\end{equation}

\noindent
And using Eq. (\ref{eq:g}) for $g$, where $G_0(x)$ is given by 
Eq. (\ref{eq:G0tri}),
we find that

\begin{equation}
g = 
\begin{dcases}
0 & \ \ \ \   p_3 \le \frac{p_1}{3} \\
1 - \frac{p_1^2}{3p_3} - \frac{p_1^2 p_2}{9 p_3^2} 
- \frac{p_1^3}{27 p_3^2} &
\ \ \ \   p_3 > \frac{p_1}{3}.
\end{dcases}
\end{equation}

\noindent
Thus, the percolation threshold is located at $p_3 = p_1/3$.
This can be understood intuitively by recalling that the finite 
components exhibit tree structures. In a tree that
includes a single node of degree $k=3$, with three chains
of arbitrary lengths attached to it, there must be
three leaf nodes of degree $k=1$. 
In more complex tree
structures, let alone in the giant 
component, there must be more than one node of
degree $3$ for every three nodes of degree $1$.
This is not likely to occur in case that $p_3 < p_1/3$.
Using the normalization condition, we find that for any given
value of $p_2$, a giant component exists for $p_3 > (1-p_2)/4$.

The degree distribution on the giant component is given by

\begin{equation}
P(k | 1) = \frac{ 1 - \left( \frac{p_1}{3 p_3} \right)^k }
{1 - \left( \frac{p_1^2}{3 p_3} \right) 
- \left( \frac{p_1^2 p_2}{9 p_3^2} \right)
- \left( \frac{p_3}{27 p_3^2} \right) } 
P(k),
\end{equation}

\noindent
where $k=1, 2, 3$ and $P(k)$ is given by Eq. (\ref{eq:trinary}).
The degree distribution on the finite components is given by

\begin{equation}
P(k | 0) = \frac{ \left( \frac{p_1}{3 p_3} \right)^k }
{ \left( \frac{p_1^2}{3 p_3} \right)
+ \left( \frac{p_1^2 p_2}{9 p_3^2} \right)
+ \left( \frac{p_3}{27 p_3^2} \right) } 
P(k).
\end{equation}

\noindent
Thus, the mean degree on the giant component is given by

\begin{equation}
\mathbb{E}[k | 1] = \frac{ 1 - \left( \frac{p_1}{3 p_3} \right)^k }
{1 - \left( \frac{p_1^2}{3 p_3} \right) 
- \left( \frac{p_1^2 p_2}{9 p_3^2} \right)
- \left( \frac{p_3}{27 p_3^2} \right) } 
\langle K \rangle,
\end{equation}

\noindent
while the mean degree on the finite components is given by

\begin{equation}
\mathbb{E}[k | 0] = \frac{ \left( \frac{p_1}{3 p_3} \right)^k }
{ \left( \frac{p_1^2}{3 p_3} \right) 
+ \left( \frac{p_1^2 p_2}{9 p_3^2} \right)
+ \left( \frac{p_3}{27 p_3^2} \right) } 
\langle K \rangle.
\end{equation}

\noindent
The assortativity coefficient of the ternary network is given by

\begin{equation}
r = -\frac{18 p_1^2 p_3^2}
{27 p_2 p_3^3 + 9 p_1^2 p_3 (p_2+2p_3) 
+27 p_1p_3^2(p_2+2p_3) + p_1^3(p_2+6p_3)}.
\end{equation}

\noindent
As $p_3$ is increased, 
while keeping $p_2$ fixed,
the network becomes denser and
the fraction of nodes, $g$, which reside on the giant component
increases, reaching $g=1$ at $p_3=1-p_2$ 
(namely, at the point in which the number of leaf nodes vanishes).
Above this point the giant component encompasses the entire
network and the assortativity coefficient $r$ vanishes.
In the opposite case, in which $p_3$ is decreased 
the network becomes more sparse.
The percolation transition takes place at
$p_{3,c}=(1-p_2)/4$.
In the limit of sparse networks just above the percolation
threshold the assortativity coefficient can be approximated by

\begin{equation}
r \simeq -3 \frac{1-p_2}{9+7p_2} 
- 64 p_2 \frac{p_3 - p_{3,c}}{(9+7p_2)^2} + {\cal{O}} \left[ (p_3 - p_{3,c})^2 \right].
\end{equation}

In the limit of $p_3 \rightarrow 1$ (and $p_1,p_2 \rightarrow 0$),
the ternary network becomes a random regular graph (RRG) with a degenerate
degree distribution of the form
$P(k) = \delta_{k,3}$,
while in the limit of 
$p_2 \rightarrow 1$ (and $p_1,p_3 \rightarrow 0$)
it becomes an RRG with
$P(k) = \delta_{k,2}$.
In general, random regular graphs
exhibit degree distributions of the form
$P(k) = \delta_{k,c}$,
where 
$c \ge 2$ is an integer.
In RRGs with $c \ge 3$
the giant component 
encompasses the entire network,
namely 
$g = \tilde g = 1$
\cite{Bonneau2017}.
Thus, the degree distribution of the giant component is simply
$P(k) = \delta_{k,c}$.

The case of an RRG with $c=2$, 
which corresponds to the limit of $p_2 \rightarrow 1$
and $p_1,p_3 \rightarrow 0$,
is special.
An RRG with $c=2$ consists of a collection of closed cycles.
The local structure of all the cycles is identical,
and follows the overall degree distribution of the network, 
$P(k) = \delta_{k,2}$.
Thus, unlike the case of other configuration model networks
there is no further information to be revealed 
about the degree distribution of the giant component. 
The generating function method used in this paper 
does not permit the calculation of the
percolating fraction, $g$, in the case of RRGs with $c = 2$, 
as the value of $g$ turns out to 
be indeterminate in this case.
However, an interesting
analogy between the cycles of RRGs with $c=2$
and the cycles which appear in they theory of random permutations, 
enables one to conclude that the average size of the longest
cycle is extensive in $N$. 
In random permutations of $N$ objects, the average length of
the longest cycle turns out to be 
$g N$ where $g \simeq 0.62455$ is the Golomb-Dickman constant
\cite{Shepp1966}.
However, in numerical simulations of RRGs with $c=2$
we found that the average length
of the longest cycle is given by $g N$, where $g \simeq 0.755$.
This difference can be understood from the fact that 
the two systems differ in some details.
For example, unlike the case
of random permutations, in RRGs with $c=2$ fixed points
(namely isolated nodes) and cycles
of length $2$ (namely dimers) are not allowed,
and the minimal cycle length is $3$.

\subsection{Configuration model networks with a Zipf degree distribution}

Consider a configuration model network with a Zipf degree distribution 
of the form

\begin{equation}
P(k) = \frac{e^{ \alpha k_{\rm min}}}{\Phi(e^{-\alpha},1,k_{\rm min})} 
\frac{e^{- \alpha k} }{k},
\end{equation}

\noindent
where $\Phi(z,s,k)$ is the Lerch transcendent function 
\cite{Olver2010}.
This distribution exhibits a power-law component of the form $k^{-\gamma}$, with
$\gamma=1$, with a cutoff in the form of an exponential tail controlled by the parameter 
$\alpha$, which sets the range of the tail.
The mean degree is given by

\begin{equation}
\langle K \rangle =
\frac{\Phi(e^{-\alpha},0,k_{\rm min}) }{\Phi(e^{-\alpha},1,k_{\rm min})}.
\end{equation}

\noindent
The generating functions take the form

\begin{equation}
G_0(x) = x^{k_{\rm min}} \Phi(x e^{-\alpha},1,k_{\rm min}),
\end{equation}

\noindent
and

\begin{equation}
G_1(x) = \frac{x^{k_{\rm min}-1}
\Phi(x e^{-\alpha},0,k_{\rm min}) }{\Phi(e^{-\alpha},0,k_{\rm min})}.
\label{eq:G1zipf}
\end{equation}

\noindent
Note that $G_0(0)=0$ and $G_0(1)=G_1(1)=1$.

From this point and on, we focus on the case $k_{min}=1$, where many of the 
quantities mentioned above become significantly simpler. 
In particular, the mean degree becomes

\begin{equation}
\langle K \rangle =
\frac{-1}{(e^{\alpha}-1)\ln (1-e^{-\alpha})},
\end{equation}

\noindent
and the two degree generating functions become

\begin{equation}
G_0(x) = \frac{\ln \left( 1-e^{-\alpha} x \right)}{\ln \left(1-e^{-\alpha}\right)},
\end{equation}

\noindent
and

\begin{equation}
G_1(x) = \frac{1-e^{-\alpha} }{1-e^{-\alpha} x}.
\label{eq:G1zipf2}
\end{equation}

\noindent
Inserting the expression of $G_1(x)$  
given by Eq. (\ref{eq:G1zipf2}) into
Eq. (\ref{eq:tg-fpe}) 
we find that there is a non-trivial solution 
of the form

\begin{equation}
\tilde g = 2 - e^{\alpha}.
\label{eq:tgzipf}
\end{equation}

\noindent
Using Eq. (\ref{eq:g}) we find that

\begin{equation}
g = 1 + \frac{\alpha}{ \ln (1-e^{-\alpha})}.
\label{eq:gzipf}
\end{equation}

\noindent
The percolation transition takes place at 
$\alpha=\ln 2$, 
below which there is a giant component.
The degree distribution 
$P(k|1)$ on the giant component is
given by 
Eq. (\ref{eq:pk1}), 
where
$\tilde g$ 
is given by Eq. 
(\ref{eq:tgzipf})
and $g$ is given by
Eq. (\ref{eq:gzipf}).
The mean degree on the giant component is given by

\begin{equation}
\mathbb{E}[K|1] =
\frac{ e^{\alpha} - 2  }
{  (1 - e^{-\alpha}) \left[ \alpha +  \ln (1-e^{-\alpha}) \right] }.
\end{equation}

\noindent
The assortativity index is given by

\begin{equation}
r = - \frac{(1-e^{-\alpha})^2}{e^{2 \alpha} -3 e^{\alpha} +3}.
\end{equation}

\noindent
For small values of $\alpha$ we obtain

\begin{equation}
r \simeq - \alpha^2 - \frac{\alpha^4}{12 } 
+ {\mathcal O} \left( \alpha^6 \right).
\end{equation}

\noindent
For values of $\alpha$ just above the percolation threshold,
we obtain

\begin{equation}
r \simeq - \frac{1}{4} + \frac{5}{4} 
 \left( \alpha - \ln 2 \right)^2
+ {\mathcal O} \left[ \left( \alpha - \ln 2 \right)^3 \right].
\end{equation}

\subsection{Configuration model networks with a power-law degree distribution}

Consider a configuration model network with a power-law degree distribution
of the form

\begin{equation}
P(k) = \frac{A}{ k^{\gamma} }
\label{eq:PLnorm1}
\end{equation}

\noindent
for $k_{\rm min} \le k \le k_{\rm max}$,
where the normalization coefficient is

\begin{equation}
A = \frac{1}{ \zeta(\gamma,k_{\rm min}) - \zeta(\gamma,k_{\rm max}+1) }
\label{eq:PLnorm2}
\end{equation}

\noindent
and $\zeta(s,a)$ is the Hurwitz zeta function 
\cite{Olver2010}.
The mean degree is given by

\begin{equation}
\langle K \rangle = 
\frac{ \zeta(\gamma-1,k_{\rm min}) - \zeta(\gamma-1,k_{\rm max}+1) }
{ \zeta(\gamma,k_{\rm min}) - \zeta(\gamma,k_{\rm max}+1) },
\label{eq:Kmsf}
\end{equation}

\noindent
while the second moment of the degree distribution is

\begin{equation}
\langle K^2 \rangle = 
\frac{ \zeta(\gamma-2,k_{\rm min}) - \zeta(\gamma-2,k_{\rm max}+1) }
{ \zeta(\gamma,k_{\rm min}) - \zeta(\gamma,k_{\rm max}+1) }.
\label{eq:K2msf}
\end{equation}

\noindent
For $\gamma \le 2$ the mean degree diverges when 
$k_{\rm max} \rightarrow \infty$.
For $2 < \gamma \le 3$ the mean degree is 
bounded while the second moment,
$\langle K^2 \rangle$, diverges.
For $\gamma > 3$ both moments are bounded.
For $\gamma > 2$ and
$k_{\rm min} \ge 2$ 
(where nodes of degrees $0$ and $1$ do not exist),
$\langle K^2 \rangle > 2 \langle K \rangle$
namely the Molloy and Reed criterion is satisfied
and the network exhibits a giant component
\cite{Molloy1995,Molloy1998}. 
Moreover, under these conditions
the giant component encompasses the entire network
\cite{Bonneau2017}.

The case of $k_{\rm min}=1$ 
is particularly interesting.
In this case, the degree distribution is given by Eq.
(\ref{eq:PLnorm1}) with 

\begin{equation}
A = \frac{1}{\zeta(\gamma)-\zeta(\gamma,k_{\rm max}+1)},
\end{equation}

\noindent
and its first two moments are

\begin{equation}
\langle K \rangle = 
\frac{\zeta(\gamma-1) - \zeta(\gamma-1,k_{\rm max}+1)}
{\zeta(\gamma) - \zeta(\gamma,k_{\rm max}+1)},
\end{equation}

\noindent
and

\begin{equation}
\langle K^2 \rangle = 
\frac{\zeta(\gamma-2) - \zeta(\gamma-2,k_{\rm max}+1)}
{\zeta(\gamma) - \zeta(\gamma,k_{\rm max}+1)}.
\end{equation}

\noindent
The generating functions of this degree distribution are

\begin{equation}
G_0(x) = \frac{ {\rm Li}_{\gamma}(x) - x^{k_{\rm max}+1} \Phi(x,\gamma,k_{\rm max}+1) }{\zeta(\gamma) - \zeta(\gamma,k_{\rm max}+1)}
\label{eq:G0sf}
\end{equation}

\noindent
and

\begin{equation}
G_1(x) = \frac{ {\rm Li}_{\gamma-1}(x)  -  x^{k_{\rm max}+1} \Phi(x,\gamma-1,k_{\rm max}+1)}
{x \left[\zeta(\gamma-1) - \zeta(\gamma-1,k_{\rm max}+1) \right]},
\label{eq:G1sf}
\end{equation}

\noindent
where ${\rm Li}_{\gamma}(x)$
is the polylogarithmic function.
To obtain a self-consistent equation for the parameter $\tilde g$, one inserts the
expression for $G_1(x)$ from Eq. (\ref{eq:G1sf})
into Eq. (\ref{eq:tg-fpe}).
\noindent
In this case, we do not have a closed form expression
for $\tilde g$, and the equation is solved numerically.
The value of $\tilde g$ is then inserted into Eq. (\ref{eq:g})
to obtain the parameter $g$.

\begin{figure}
\begin{center}
\includegraphics[width=0.45\textwidth]{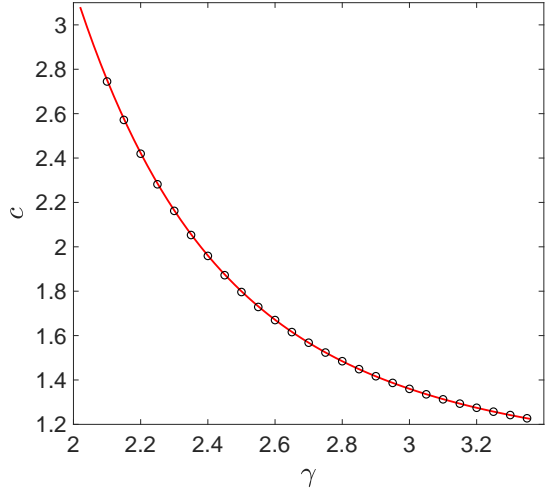} 
\end{center}
\caption{
(Color online)
The mean degree, 
$c= \langle K \rangle$,
as a function of the exponent $\gamma$,
for a configuration model network 
with a power-law degree
distribution, where $k_{\rm min}=1$ and $k_{\rm max}=100$.
As $\gamma$ is increased, the tail of the degree distribution 
decays more quickly and as a results the mean degree decreases.
The analytical results (solid line)
are found to be in excellent agreement with the results of computer 
simulations (circles) performed for networks
of $N=4 \times 10^4$ nodes.
}
\label{fig:5}
\end{figure}

In Fig. 
\ref{fig:5} 
we present analytical results for the mean degree
$c=\langle K \rangle$
(solid line),
of a configuration model network with a power-law degree
distribution and $k_{\rm min}=1$,
as a function of the exponent $\gamma$,
for $\gamma > 2$.
As $\gamma$ is increased, the mean degree, $c$ decreases.
The analytical results are in excellent agreement with the results
of computer simulations (circles).

\begin{figure}
\begin{center}
\includegraphics[width=0.45\textwidth]{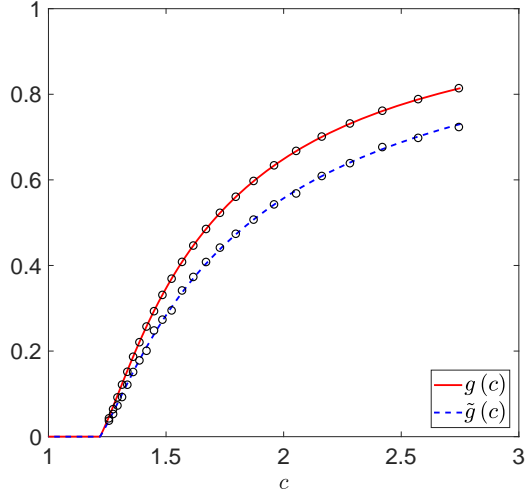} 
\end{center}
\caption{
(Color online)
Analytical results for the parameters
$g$ (solid line) 
and
$\tilde g$ 
(dashed line),
as a function of the mean degree,
$c$,
for a configuration model network 
with a power-law degree
distribution, where $k_{\rm min}=1$ and $k_{\rm max}=100$.
The analytical results 
are found to be in excellent agreement with the results of computer 
simulations (circles), performed for networks 
of size
$N=4 \times 10^4$.
}
\label{fig:6}
\end{figure}

In Fig.
\ref{fig:6}
we present analytical results for the parameters
$g$ (solid line)
and
$\tilde g$
(dashed line)
for a configuration model network with a power-law degree
distribution and $k_{\rm min}=1$,
as a function of the mean degree, $c$.
Below the percolation threshold there is no giant component
and thus $g = \tilde g =0$.
Above the percolation threshold both $g$ and $\tilde g$
gradually increase towards the dense limit result
of $g = \tilde g =1$.

\begin{figure}
\begin{center}
\includegraphics[width=0.45\textwidth]{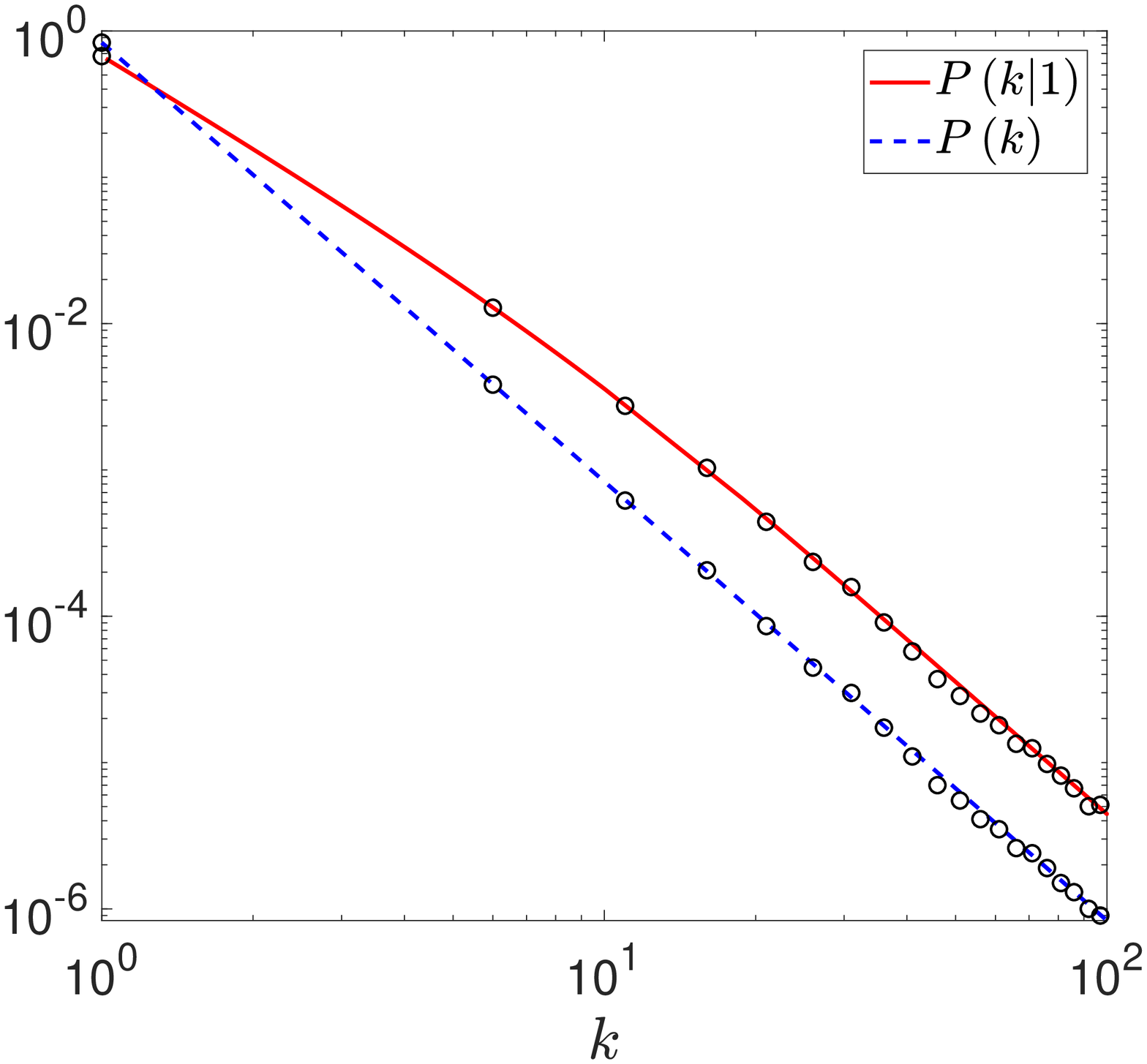} 
\end{center}
\caption{
(Color online)
Analytical results for the
degree distribution
$P(k|1)$
of the giant component of a configuration model
network with a power-law degree distribution,
where
$\gamma=3$,
$k_{\rm min}=1$
and $k_{\rm max}=100$
(solid line),
obtained from Eq. (\ref{eq:pk1}),
where $\tilde g$ and $g$ are obtained
using the generating functions of Eqs. (\ref{eq:G0sf}) and (\ref{eq:G1sf}).
The degree distribution
$P(k)$ 
of the whole network
(dashed line),
obtained from Eq. (\ref{eq:PLnorm1}),
is also shown for comparison.
The analytical results 
are found to be in excellent agreement with the results of computer 
simulations (circles), performed for networks
of $N=4 \times 10^4$ nodes,.
}
\label{fig:7}
\end{figure}

In Fig. \ref{fig:7} we present analytical results for the degree distribution
$P(k|1)$ (solid line)
of the giant component of a configuration model network with
a power-law degree distribution
where $k_{\rm min}=1$ and $\gamma=3$.
For comparison, we also show the degree distribution $P(k)$ of the whole
network (dashed line).
The analytical results are in excellent agreement with the results of computer simulations
(circles).
Using Eq. (\ref{eq:c_1}), we calculate the mean degree
on the giant component and obtain

\begin{equation}
\mathbb{E}[K | 1] = 
\frac{\tilde g (2 - \tilde g) \left[ \zeta(\gamma-1) - \zeta(\gamma-1,k_{\rm max}+1) \right] }
{ g \left[ \zeta(\gamma) - \zeta(\gamma,k_{\rm max}+1) \right] }.
\label{eq:Ek1}
\end{equation}

\begin{figure}
\begin{center}
\includegraphics[width=0.45\textwidth]{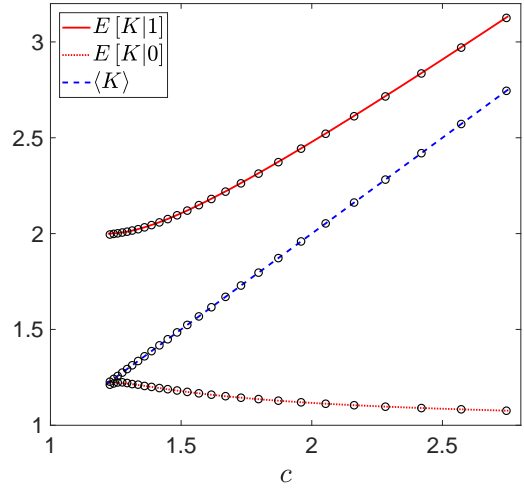} 
\end{center}
\caption{
(Color online)
Analytical results for the
mean degree
$\mathbb{E}[K|1]$
of the giant component (solid line)
and the mean degree
$\mathbb{E}[K|0]$
of the finite components (dotted line),
as a function of $c$ above the percolation threshold,
for a scale free configuration model
network, where $k_{\rm min}=1$ and $k_{\rm max}=100$. 
The analytical results 
are found to be in excellent agreement with the results of computer 
simulations (circles), performed for networks 
of $N=4 \times 10^4$ nodes.
The mean degree
$\langle K \rangle = c$ 
of the whole network
(dashed line)
is also shown for comparison.
}
\label{fig:8}
\end{figure}

\noindent
In Fig. \ref{fig:8} we present analytical results for the mean degree,
$\mathbb{E}[K|1]$, 
of the giant component 
(solid line)
and the mean degree
$\mathbb{E}[K|0]$,
of the finite components
(dotted line).
The mean degree
$\langle K \rangle$ 
of the whole network
(dashed line)
is also shown for comparison.
The analytical results are in excellent agreement with the results of
computer simulations.

Inserting the expression for
$G_1(x)$ from
Eq. (\ref{eq:G1sf}) into Eqs. (\ref{eq:E2}) and (\ref{eq:S}) 
we obtain the functions
$B(u,v)$ and $S(u)$, respectively.
Inserting them into Eq. (\ref{eq:R}) we obtain the
assortativity coefficient $r$.

\begin{figure}
\begin{center}
\includegraphics[width=0.45\textwidth]{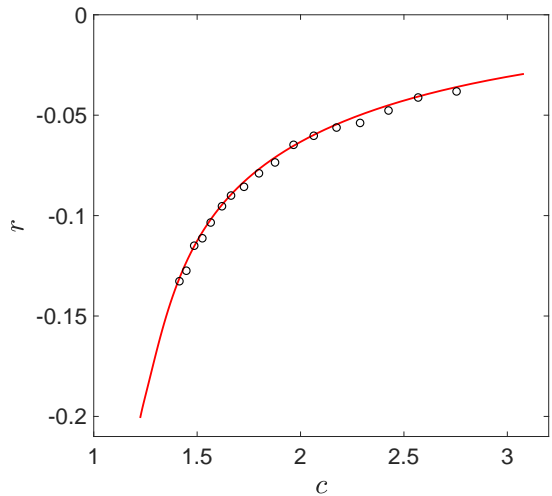} 
\end{center}
\caption{
(Color online)
The assortativity coefficient $r$ of the giant component of a configuration model
network with a power-law degree distribution,
with $k_{\rm min}=1$ and and $k_{\rm max}=100$,
as a function of the mean degree, $c$.
The analytical results (solid line)
are found to be in very good agreement with the results of computer 
simulations (circles), performed for networks
of $N=8 \times 10^4$ nodes.
The negative values of $r$ imply that the giant component is disassortative,
namely high degree nodes preferentially connect to low degree nodes
and vice versa.
As $c$ is increased, the giant component encompasses an increasing fraction
of the entire network, and $r$ decays to zero.
}
\label{fig:9}
\end{figure}

In Fig. \ref{fig:9} we present analytical results (solid line) for the assortativity coefficient
$r$ of the giant component of a configuration model network with
a power-law degree distribution,
as a function of the mean degree, $c$.
The analytical results are found to be in very good agreement with
the results of computer simulations.
For small values of $c$, just above the percolation threshold, the coefficient
$r$ is large and negative. In this regime, the giant component encompasses only
a small fraction of the network and is highly correlated. As $c$ is increased, the 
size of the giant component increases 
encompassing a larger fraction of the nodes in the network,
and the assortativity coefficient gradually decays to zero.
Using entropic considerations, one can show that negative 
degree-degree correlations are indeed typical in scale-free networks
\cite{Johnson2010,Williams2014}.

\section{Percolation on the giant component}
\label{Sec:PercGC}

Consider an ensemble of random networks of size $N$, 
with a given degree statistics,
which can be characterized by a degree distribution, $P(k)$,
degree-degree correlations and possibly higher order correlations.
The probability of a random
node to reside on the giant component is $g$,
while the probability of a random neighbor
of a random node to reside on the giant component is $\tilde g$.
Thus, in the limit of $N \rightarrow \infty$, the size of the
giant component is $gN$. 
Here we focus on the sub-network which consists of the giant
component of the primary network.
Clearly, this network consists of a single connected component.
This property is reflected in the fact that taking
the complete degree statistics of the primary network model 
and conditioning on its giant component, 
the probability, $\sigma$, that a random node will reside on the giant component 
must satisfy $\sigma=1$.

In what follows, we use generating functions to explore how 
well the result of $\sigma=1$ 
is reproduced, when degree statistics 
conditioned on the giant component is used in approximate ways. 
Following the classical percolation theory, as summarized by Eq. (\ref{eq:g}), 
the probability $\sigma$ satisfies

\begin{equation}
\sigma =  \sum_k P(k|1) \left[1 - (1- \tilde \sigma)^k \right],
\end{equation}

\noindent
where $P(k|1)$ is the degree distribution conditioned on 
the giant component given by Eq. (\ref{eq:pk1}), 
and $\tilde \sigma$ 
is the probability that a randomly chosen edge points to a 
node connected to the giant component. 
Hence

\begin{equation}
\sigma =  \frac{1}{g} \sum_k P(k) 
\left[1 - (1- \tilde g)^k \right] 
\left[1 - (1- \tilde \sigma)^k \right],
\end{equation}

\noindent
which can also be written in the form

\begin{equation}
\sigma
=  \frac{1}{g} \left[1 - G_0(1- \tilde g)
- G_0(1- \tilde \sigma)+G_0((1- \tilde g)(1- \tilde \sigma))\right].
\end{equation}

\noindent
In order to utilize these equations, one should first
calculate $\tilde \sigma$.
This is done using an approximate self-consistency equation
for $\tilde \sigma$. 
One can derive several variants for this equation, 
which depend on the level of detail in which the 
degree-degree correlations 
are taken into account.
Below we present two such variants. 
In the first variant we account only for the degree distribution,
$P(k|1)$, ignoring the degree-degree correlations. 
This variant resembles the self-consistency equation [Eq. (\ref{eq:tg-fpe})]. 
In the second variant we account for 
both the degree distribution and the
correlations between the degrees of adjacent nodes.

\subsection{Configuration Model Approximation} 

We first consider the simplest approximation, in which the 
degree-degree correlations are ignored. In this case, 
the giant component is considered as a configuration model 
network with the degree distribution $P(k|1)$. 
In this approximation 
the self-consistency equation for 
$\tilde \sigma$ 
is given by

\begin{equation}
1-\tilde \sigma 
= 
\sum_k \frac{k}{c_1} P(k|1) (1-\tilde \sigma)^{k-1},
\end{equation}

\noindent
where $P(k|1)$ given by Eq. (\ref{eq:pk1}) and $c_1$ 
is given by Eq. (\ref{eq:c_1}). 
This equation reflects the same reasoning as in Eq. (\ref{eq:tg-fpe}).
Inserting $P(k|1)$ and expressing the right hand side in
terms of the generating functions, one obtains

\begin{equation}
1 - \tilde \sigma = \frac{1}{1-(1-\tilde g)^2} 
\left[G_1(1-\tilde \sigma)-(1-\tilde g)
G_1((1-\tilde g)(1-\tilde \sigma)) \right].
\end{equation}

\subsection{Approximation using degree-degree correlations}

Taking degree-degree correlations into account as 
encoded in the degree distribution, 
$\widetilde P(k|1)$,
of random neighbors of random nodes, 
given by 
Eq. (\ref{eq:p1k}), 
one obtains a self-consistency
equation 
of the form

\begin{equation}
1-\tilde \sigma = \sum_k  \widetilde P(k|1) (1-\tilde \sigma)^{k-1},
\end{equation}

\noindent
or

\begin{equation}
1 - \tilde \sigma =
\frac{1}{g} \left\{(1-p_0)G_1(1-\tilde \rho) 
- \frac{1-g-p_0}{1-\tilde g}
G_1[(1-\tilde g)(1-\tilde \sigma)] \right\}.
\end{equation}

\begin{figure}
\begin{center}
\includegraphics[width=0.45\textwidth]{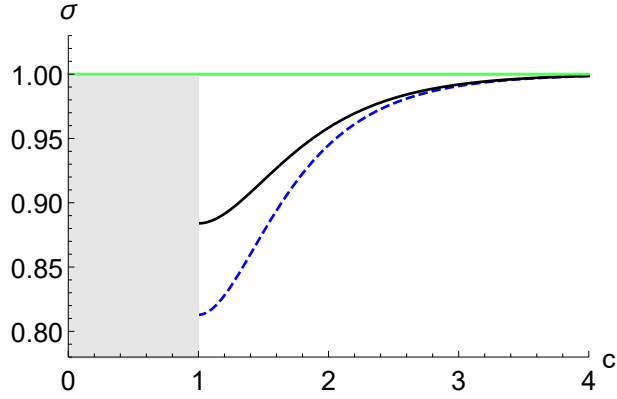} 
\end{center}
\caption{
(Color online)
The probability, $\sigma$, that a random node on the giant component of an ER network,
will remain on the giant component in 
approximate self-consistent formulations,
as a function of $c$.  
The results obtained by taking into account only the degree distribution, $P(k|1)$, 
are shown by a dashed line (blue), while the results obtained by adding the 
degree-degree correlations are shown by a solid line (black).
It is found that the
inclusion of degree-degree correlations significantly improves the results,
bringing them closer to the exact result of $\sigma=1$ for $c > 1$.
}
\label{fig:10}
\end{figure}

\noindent
In Fig. \ref{fig:10} we present the
fraction, $\sigma$, of nodes on the giant component of an ER network
which are accounted for as giant component nodes
by an approximate self-consistent approach,
as a function of the mean degree $c$.
The results were obtained from a simple self-consistent
approach which takes into account only the degree distribution $P(k|1)$ 
(dashed line), and a more complete approach which includes
both the degree distribution and 
degree-degree correlations (solid line).
The inclusion of degree-degree correlations significantly improves the results,
bringing them closer to the exact result of $\sigma=1$ for $c > 1$.
However, even with these correlation included the probability $\sigma$ is still determined to be lower than $1$; the discrepancy is largest at small $c$, though never larger than $12\%$. Our results imply that additional correlations play a role in keeping the giant component as a single connected component.

However, for small values of $c$ the probability $\sigma$ is still lower than $1$,
which means that additional correlations play a role in keeping the
giant component as a connected component.

\section{Applications}
\label{Sec:ApplGC}

In what follows, we present several results that exploit 
the degree distributions and the joint degree distributions 
of higher orders obtained in 
Sec. \ref{Sec:GC}.
This analysis elucidates 
both the power and limitations of this approach.

\subsection{Distribution of shortest path lengths on the giant component} 

Consider a random node, $i$, in an ER network of $N$ nodes
and mean degree $c=(N-1)p$. The remaining $N-1$ nodes are
organized in shells, such that the $\ell$th shell consists of the nodes
which are at a distance $\ell$ from the central node, $i$.
The number of nodes in the $\ell$th shell is denoted by $n_{\ell}$,
where $n_{0}=1$.
The total number of nodes in the $(\ell+1)$th shell and all the outer shells
beyond it is given by
$N_{\ell} = \sum_{\ell'=\ell+1}^{\infty} n_{\ell}$.
Therefore, the number of nodes in the $\ell$th shell can be
expressed by
$n_{\ell} = N_{\ell-1} - N_{\ell}$.
The approach we now present for the calculation of the DSPL is called the
{\it random shells approach} (RSA)
\cite{Katzav2015}.
Within this approach $n_{\ell}$ satisfies the recursion equation
$n_{\ell+1} = N_{\ell-1} [1 - (1-p)^{n_{\ell}}]$, 
which can also be written in the form
$N_{\ell+1} = N_{\ell} (1-p)^{N_{\ell-1} - N_{\ell}}$,
where $N_0 = N-1$
and 
$N_1 = (N-1)(1-p)$.
We denote the probability that the shortest path length from node $i$
to another random node in the network is larger than $\ell$ by $P(L>\ell)$.
This probability is given by
$P(L>\ell) = N_{\ell}/(N-1)$.
Using this relation one obtains a
recursion equation for the distribution of shortest path lengths (DSPL),
which is expressed in the form of a tail distribution.
It is given by

\begin{equation}
P(L>\ell+1) = P(L>\ell) (1-p)^{(N-1)[P(L>\ell-1)-P(L>\ell)]},
\label{eq:RSA}
\end{equation}

\noindent
where
$P(L>0)=1$ 
and 
$P(L>1)=1-p$.
It is worth mentioning that the DSPL of 
an ER network can also be obtained using
the {\it random paths approach},
which is based on the shortest paths between pairs of nodes
rather than the shells around a single node
\cite{Katzav2015}.
The latter approach was extended to the case of configuration model networks
\cite{Nitzan2016,Melnik2016}.

In the following, we will make an attempt at improving this approach
based on the results derived for the giant component. 
In the classical RSA theory the reference 
node $i$ is considered as a 'typical' node in 
the spirit of mean-field theory, and its degree is assumed to be equal
to the mean degree, $c$. In practice, the degree of a random reference
node is drawn from the degree distribution $P(k)$. Moreover, the degree
of the reference node, $i$, has a strong effect on the shell structure around it.
To account for this effect, we consider the shell structure around a node $i$
of a given degree $k_0$.
In this case, 
$P(L=1|k_0) = k_0/(N-1)$ 
and 
$P(L>1|k_0)=1 - k_0/(N-1)$.
The recursion equations take the form

\begin{equation}
P(L>\ell+1 | k_0) = P(L>\ell | k_0) \left( 1 - \frac{c}{N-1} \right)^{(N-1)P(L=\ell | k_0)},
\end{equation}

\noindent
and
\begin{equation}
P(L=\ell+1 | k_0) = P(L>\ell | k_0) - P(L>\ell+1 | k_0).
\end{equation}

\noindent
Note that in case that $i$ is an isolated node, namely $k_0=0$,
one obtains 
$P(L>\ell | k_0=0) = 1$
for any values of $\ell$.
The DSPL is then assembled from these conditional probabilities, with
suitable weights, according to

\begin{equation}
P(L>\ell) = \sum_{k_0=0}^{\infty} P(k_0) P(L > \ell | k_0)
\end{equation}

\noindent
where 

\begin{equation}
P(k_0) = \frac{e^{-c} c^{k_0}}{k_0!}
\end{equation}

\noindent
is the Poisson distribution.
Separating the case of $k_0=0$ we obtain

\begin{equation}
P(L > \ell) = e^{-c} 
+ \sum_{k_0=1}^{\infty}  
\frac{e^{-c} c^{k_0}}{k_0!} P(L>\ell | k_0).
\label{eq:kRSA}
\end{equation}

\noindent
Since the analysis leading to 
Eq. (\ref{eq:kRSA}) takes into account the effect of the degree, $k_0$,
of the reference node, $i$, on the shell structure around it, 
this approach is referred to as the kRSA approach.
We will now focus on the asymptotic tail of $P(L>\ell)$, which accounts for the
fraction of nodes which are infinitely far away from $i$. On a finite network, the
asymptotic value is given by $P(L>N-1)$. In this analysis we need to distinguish
between the case in which $i$ resides on the giant component and the case in which
it resides on one of the finite components.
In case that $i$ is chosen randomly, without conditioning on its degree,
the probability that it resides on the giant component is
$P(\Lambda=1) = g$ and the probability that it resides on one of the finite 
components is $P (\Lambda=0) = 1-g$.
The degree distribution of nodes on the giant component is given by Eq. (\ref{eq:PkER}), 
while the degree distribution of nodes which reside on one of the finite components is

\begin{equation}
P(k | 0) = e^{-c(1-g)} \frac{[c(1-g)]^k}{k!}.
\end{equation}

\noindent
For a node that resides on one of the finite components, we can
approximate the DSPL by
$P(L>\ell | \Lambda = 0)=1$ 
for all values of $\ell>0$.
This is an excellent approximation because the vast majority
of pairs of nodes which are not on the giant component are not connected,
since they do not belong to the same component at all. 
More precisely, the probability that they are connected scales as $1/N$ and hence 
negligible in the large network size limit.
Under this assumption

\begin{equation}
P(L>\ell) = P(\Lambda =0) P(L>\ell | \Lambda =0)
+ P(\Lambda =1) \sum_{k_0=0}^{\infty} P(k_0 | 1) P(L>\ell |k_0),
\end{equation}

\noindent
or more explicitly

\begin{equation}
P(L > \ell) = (1-g) + 
\sum_{k_0=0}^{\infty} 
\frac{e^{-c} c^{k_0}}{k_0!} 
\left[ 1 - e^{cg} (1-g)^{k_0+1} \right] P(L>\ell | k_0).
\label{eq:kgRSA}
\end{equation}

\noindent
The analysis leading to Eq. (\ref{eq:kgRSA}) takes into account the distinction
between reference nodes which reside on the giant component (with probability $g$)
or on the non-giant components (with probability $1-g$),
respectively.
This analysis is thus referred to as the kgRSA approach.

To obtain the asymptotic value 
$P_{\infty} = P(d>N-1)$,
we insert in 
Eq. (\ref{eq:kgRSA})
the identity
$P(L>N-1 | k_0)=1-g$
for all values of $k_0$, since the $(1-g)N$ nodes which 
are not on the giant component are always beyond reach.
In this case, the sum over the degree distribution in
Eq. (\ref{eq:kgRSA}) 
is equal to $g$.
Therefore,

\begin{equation}
P_{\infty} = (1-g) + g(1-g) = 1-g^2,
\end{equation}

\noindent
which coincides with the known exact result,
namely with the probability that two random nodes 
do not reside simultaneously on the giant component.

\begin{figure}
\begin{center}
\includegraphics[width=0.3\textwidth]{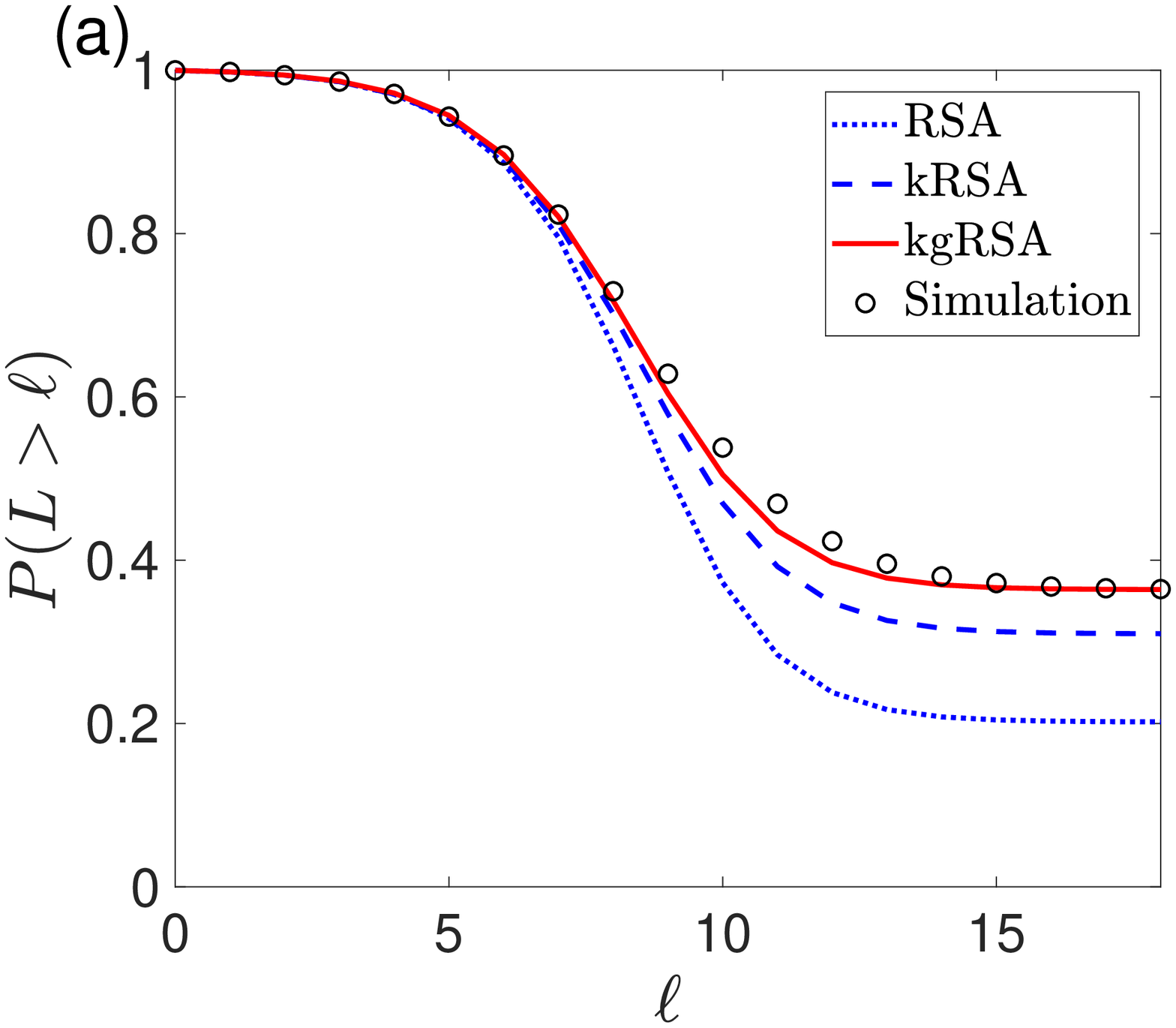} \hfil
\includegraphics[width=0.3\textwidth]{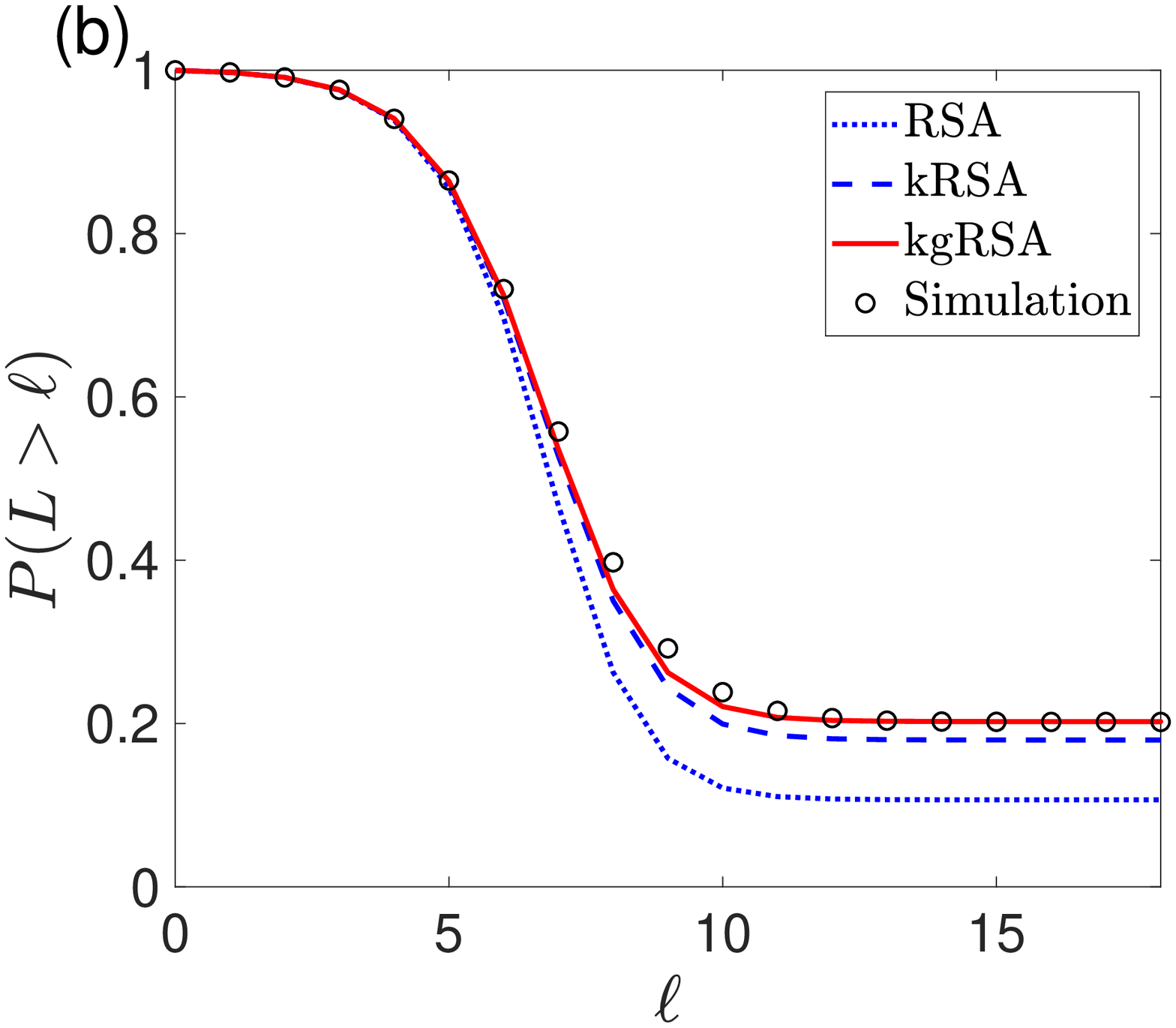} \hfil
\includegraphics[width=0.3\textwidth]{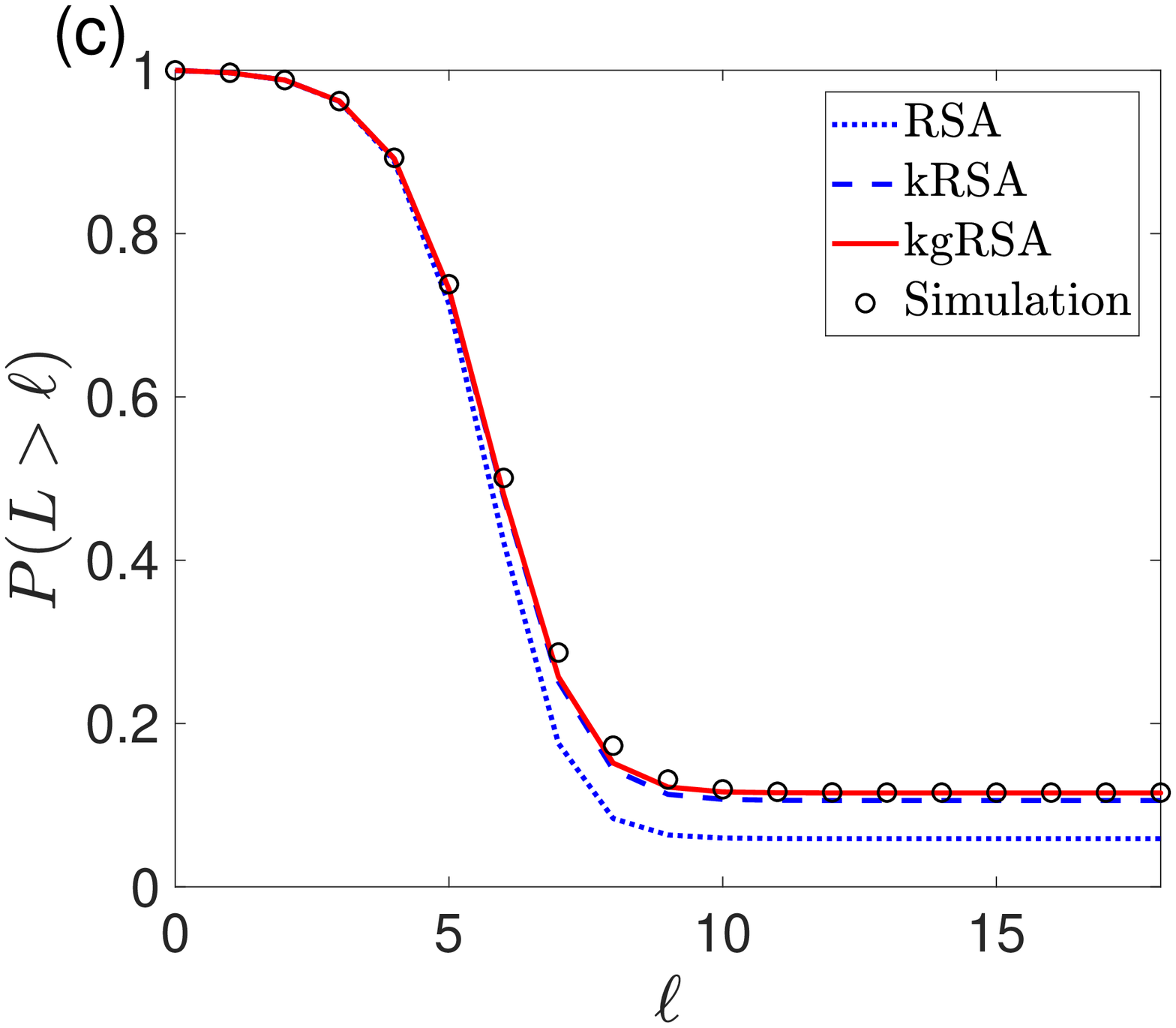}
\caption{
(Color online)
Analytical results 
(dotted, dashed and solid lines) and simulation results
(circles) for the tail distribution of the shortest path lengths,
$P(L>\ell)$, of an ER network of $N=1000$ nodes with $c=2$ (a), 
$2.5$ (b) and $3$ (c).
The RSA results (dotted lines), obtained from 
Eq. (\ref{eq:RSA}), 
are accurate for small distances but greatly under-estimate
the tail distribution for large distances.
The kRSA results (dashed lines), 
obtained from Eq. (\ref{eq:kRSA}),
which applies the overall degree distribution $P(k)$,
provide a significant improvement, but are still too low.
The kgRSA results (solid lines), 
obtained from Eq. (\ref{eq:kgRSA}),
which takes into account the degree distribution on the
giant component, $P(k|1)$,
are found to be in very good agreement 
with the simulation results
(circles).  
}
\label{fig:11}
\end{center}
\end{figure}

In Fig. \ref{fig:11} we present analytical results 
(dotted, dashed and solid lines) and simulation results
(circles) for the tail distribution
$P(L>\ell)$ of an ER network of $N=1000$ nodes with $c=2$ (a), 
$2.5$ (b) and $3$ (c).
The RSA results (dotted lines), obtained from 
Eq. (\ref{eq:RSA}), 
are accurate for small distances but greatly under-estimate
the tail distribution for large distances.
The kRSA results (dashed lines), 
obtained from Eq. (\ref{eq:kRSA}),
account for the degree
of the central node and 
whether this node belongs (or not) to the giant component.
This approach provides 
a significant improvement, but the resulting probabilities are 
still lower than the results of the simulations.
The kgRSA results (solid lines), 
obtained from Eq. (\ref{eq:kgRSA}),
account for the degree of the central node and
are summed up using $P(k|1)$.
These results are found to be in very good agreement 
with the simulation results
(circles), and coincide with the tail exactly.  

\subsection{The spectra of the giant component}

In this section we present an example of a problem 
in which the knowledge of the
degree distributions on the giant component, 
$P(k|1)$ and $\widetilde P(k|1)$,
enables to utilize a known formalism
developed for the entire network, 
for the analysis of the giant component alone.
There are many problems in which this approach
can be applied, by
replacing the degree distributions of the entire network,
$P(k)$ and $\widetilde P(k)$,
by the corresponding degree distributions conditioned
on the giant component,
$P(k|1)$ and $\widetilde P(k|1)$,
in the equations which provide the desired properties.

The specific example we consider involves the 
the calculation of the spectra of the 
adjacency matrices of configuration model networks,
using methods of random matrix theory
\cite{Mehta2004,Livan2018}.
The methodology for studying the spectrum of the adjacency matrix, 
$A$, of an entire network was developed in Refs. 
\cite{Kuhn2008,Rogers2008}.
It is based on 
a representation of the spectral density 
of a matrix $A$ in terms of the trace of 
its resolvent

\begin{equation}
\rho(\lambda) = \frac{1}{\pi N}
\lim_{\epsilon \rightarrow 0^+} {\rm Im} 
{\rm Tr} (\lambda_{\epsilon} I - A)^{-1},
\label{eq:EJ}
\end{equation}

\noindent
where $I$ is the identity matrix and
$\lambda_{\epsilon} = \lambda - i \epsilon$.
In fact, Eq. (\ref{eq:EJ}) is an example of the
Stieltjes-Peron inversion formula
\cite{Akhiezer1965}.
Edwards and Jones 
\cite{Edwards1976}
expressed the trace of the resolvent in terms of
a sum over single-site variances

\begin{equation}
\rho(\lambda) = \frac{1}{\pi N} {\rm Re}
\sum_i \langle u_i^2 \rangle,
\end{equation}

\noindent
of the complex Gaussian measure

\begin{equation}
p({\bf u}) = \frac{1}{Z}\, {\rm e}^{{\rm i} H({\bf u})}
\label{CGauss}
\end{equation}

\noindent
in which $H({\bf u})$ is given by the quadratic form

\begin{equation}
H({\bf u}) = \frac{1}{2}
\sum_{i,j} (\lambda_{\epsilon} \delta_{ij} - A_{ij}) u_i u_j.
\end{equation}

\noindent
The formalism of Refs.
\cite{Kuhn2008,Rogers2008,Kuhn2016}
enables one to express the ensemble average of
$\rho(\lambda)$ in terms of the distribution $\pi(\omega)$ of so-called inverse single-cavity variances corresponding to the multi-variate complex Gaussian (\ref{CGauss}).
It takes the form

\begin{equation}
\rho(\lambda) =
\frac{1}{\pi} {\rm Re} \sum_{k \ge 0}
P(k) \int \prod_{\nu = 1}^k d\pi(\omega_{\nu})
\frac{1}{i \lambda_{\epsilon}
+ \sum_{\nu=1}^k \frac{1}{\omega_{\nu}} }.
\label{eq:spect1}
\end{equation}

\noindent
The distribution $\pi(\omega)$ of inverse single-cavity variances is determined as solution of the
self-consistency equation

\begin{equation}
\pi(\omega) =
\sum_{k \ge 1} \widetilde P(k) \int \prod_{\nu=1}^{k-1}
d \pi(\omega_{\nu})
\delta\left(\omega - i \lambda_{\epsilon}
- \sum_{\nu=1}^{k-1} \frac{1}{\omega_{\nu}}\right).
\label{eq:spect2}
\end{equation}

\begin{figure}
\begin{center}
\includegraphics[width=0.4\textwidth]{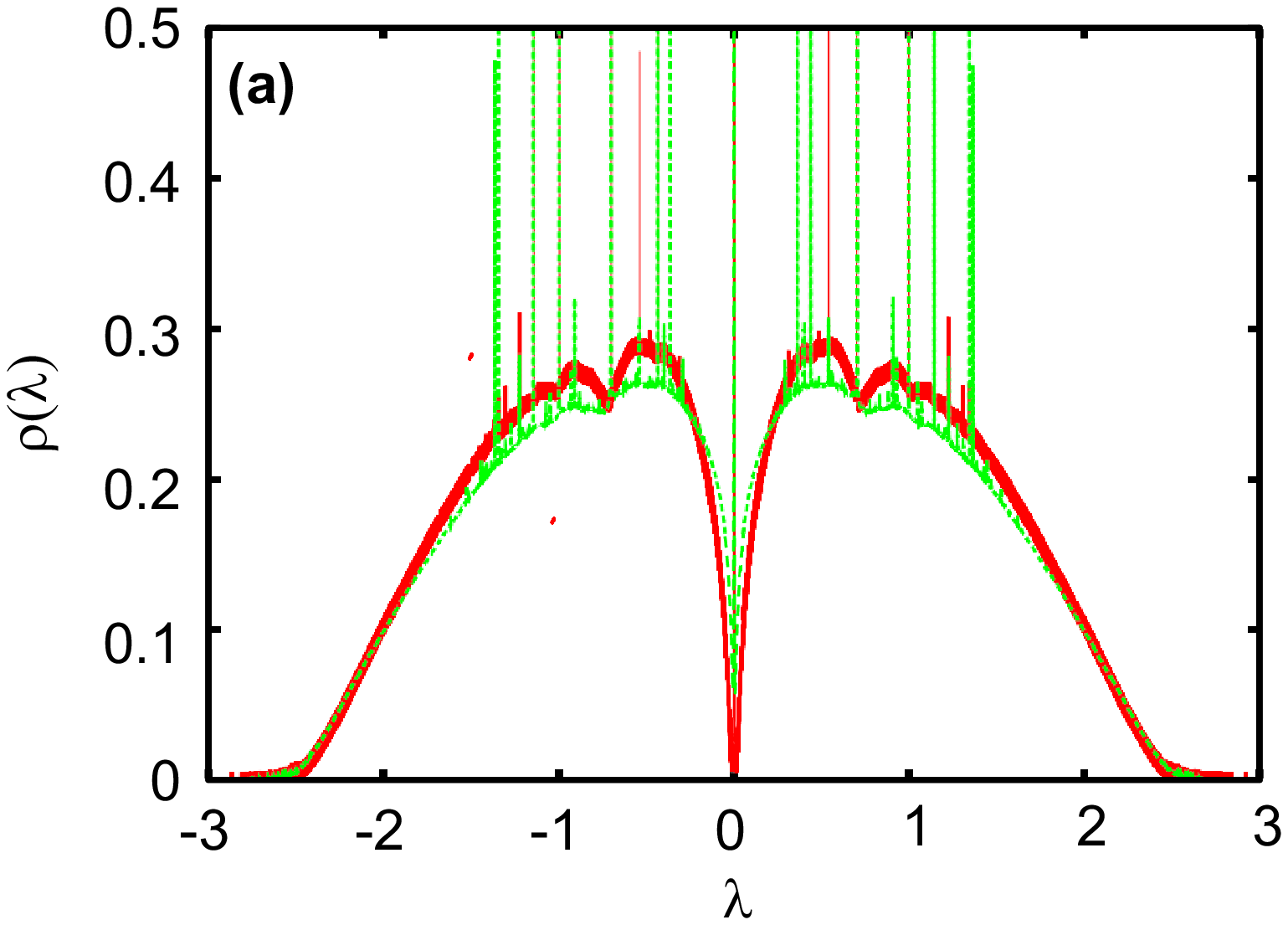} \hfil
\includegraphics[width=0.4\textwidth]{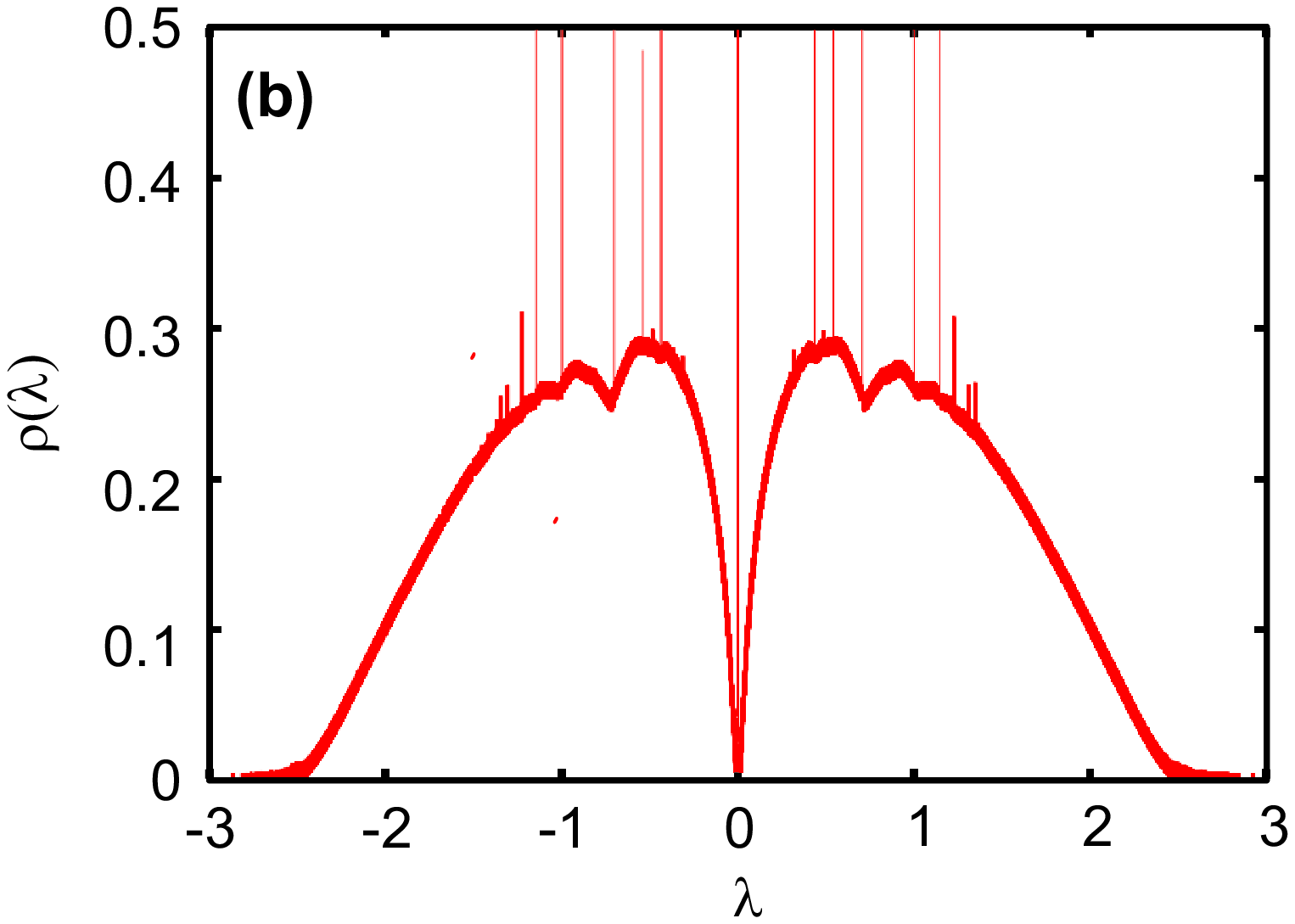}
\end{center}
\caption{
(Color online)
(a) The spectrum of the adjacency matrix of the 
giant component of an ER network of mean degree $c=2$ 
(red solid line), compared with an approximation that 
takes degree-degree correlations at the level of 
Eq. (\ref{eq:Pkk1'}) into account (green dashed line). 
The approximate description does not exclude the 
existence of finite components (as can be seen in Fig. \ref{fig:10}), so has more weight in 
localized states with support on finite components 
(represented by $\delta$-peaks in the spectrum); 
(b) The exact spectrum shown separately, demonstrating 
that the exact spectrum of the giant component also 
exhibits a number of localized states (corresponding 
to sub-graph configurations with 
$\mathbb{Z}_2$ symmetry).
}
\label{fig:12} 
\end{figure}

In this work we adapt the ensemble averaging step in
the formalism reviewed above to reflect the degree 
distributions and degree-degree correlations on the giant component.
This amounts to replacing $P(k)$ in Eq. (\ref{eq:spect1}) by
$P(k|1)$ and replacing $\widetilde P(k)$ in Eq. (\ref{eq:spect2})
by $\widetilde P(k|1)$ [Eq. (\ref{eq:p1k})].
We apply this approach to the calculation of the 
spectrum of the adjacency matrices, 
conditioned on the giant component, for an ensemble of
ER networks with $c=2$ in the large $N$ limit.
In Fig. \ref{fig:12} we present the
resulting spectrum 
(solid line), compared with an approximation that 
takes degree-degree correlations at the level of 
Eq. (\ref{eq:p1k}) 
into account (dashed line). 
The approximate description does not exclude the 
existence of finite components (as seen in section VII B), so has more weight in 
localized states with support on finite components 
(represented by $\delta$-peaks). 
We also show separately the
exact spectrum calculated using the approach of Ref. 
\cite{Kuhn2016}, 
demonstrating that the exact spectrum of the giant component also 
exhibits a number of localized states (corresponding 
to sub-graph configurations with $\mathbb{Z}_2$ symmetry).

\subsection{Epidemic spreading on the giant component} 

One of the most important dynamical processes taking place
on networks is the spreading or propagation of 
infections, information and opinions.
To discuss processes of epidemic spreading on a network,
let us first define the possible states of a node as
susceptible (S), infected (I) or recovered (R) \cite{Karrer2010,Rogers2015,Satorras2015}.
The transitions between these states include, for example, S $\rightarrow$ I,
where a susceptible node becomes infected due to the
interaction with an infected neighbor. In the 
susceptible-infected-susceptible (SIS) model, the infected
node later recovers and returns to the S state, while
in the susceptible-infected-recovered (SIR) model, the
infected node recovers and becomes immune to further 
infections. The classical epidemic models have been studied 
extensively leading to many insights 
and applications such as assessment of vaccination strategies.

The infection is considered as a stochastic process,
starting from a random infected node, and propagates
through the network. Each infected node infects 
each of its neighbors with probability $\rho$.
In each instance of this process, the number of infected
nodes exhibits temporal fluctuations until the infection
dies out.
The statistical properties of the infections depend on
the network structure and on the parameter $\rho$. 
The long term dynamics of
an SIR model on a network can be mapped into a bond 
percolation problem on the network \cite{Karrer2010}.
The percolation problem involves a random deletion 
of edges, such that each edge in the network is
maintained with probability $\rho$ and deleted with
probability $1-\rho$.
Within this construction,
the probability, $\sigma$, of a random node 
in a configuration model network to remain on
the giant component corresponds to the 
fraction of the individuals which have become infected. 
This fraction is given by 

\begin{equation}
\sigma = 1 -\sum_{k=0}^{\infty} P(k) (1 - \rho \tilde \sigma)^k 
= 1 - G_0(1-\rho \tilde \sigma),
\end{equation}

\noindent
where $\tilde \sigma$ is the probability that a 
random neighbour of a random node is a part of the
giant component,
which satisfies the self-consistency equation

\begin{equation}
\tilde \sigma = 1 - \sum_k \frac{k}{c} P(k)
(1-\rho\tilde \sigma)^{k-1}
=
 1 - G_1(1-\rho\tilde \sigma).
\label{eq:tg2}
\end{equation}

\noindent
These equations closely resemble those used to 
identify the fraction of nodes in the giant component 
of the primary network discussed above in section VII. 
We can now apply some of the heuristics 
developed in this work to recover 
aspects of heterogeneity in the percolation problem 
on random networks that were recently described in 
\cite{Kuhn2017},
{\it without} having to apply the message passing 
and population dynamics techniques used in that paper.

As in Sec. \ref{Sec:PercGC}  we rewrite the equation for 
$\sigma$ in a manner that allows us to explore its 
probabilistic content, by iteratively inserting the 
self-consistency equation (\ref{eq:tg2}) for
$\tilde \sigma$. 
In order to achieve more compact versions 
for the resulting expression we choose to rewrite 
Eq. (\ref{eq:tg2}) 
as an equation for the variable 
$\tilde h \equiv 1-\rho \tilde \sigma$, 
giving 

\begin{equation}
\tilde h 
= 1 - \rho \left[1 - G_1(\tilde h) \right]
= \sum_k \frac{k}{c} P(k) \left[ 1-\rho(1- \tilde h^{k-1}) \right].
\label{eq:th}
\end{equation}

\noindent
To first order, we obtain

\begin{equation}
\sigma = \sum_{k=0}^{\infty} P(k)  \left( 1 - \tilde h^k \right).
\label{eq:g0}
\end{equation}

\noindent
Replacing $\tilde h$ in Eq. (\ref{eq:g0}) 
by the right hand side of Eq. (\ref{eq:th}),
we obtain  

\begin{equation}
\sigma = 
\sum_{k;\{k_\mu\}} 
P(k) \prod_{\mu=1}^k \frac{k_\mu}{c} P(k_\mu)
\left[1 - \prod_{\mu=1}^k \left(1-\rho (1-\tilde h^{k_\mu-1}) \right) \right].
\label{eq:g1}
\end{equation}

\noindent
Repeating this procedure once again we obtain

\begin{eqnarray}
\sigma    &=& \sum_{k;\{k_\mu\};\{k_{\mu\nu}\}} 
P(k) \prod_{\mu=1}^k
\frac{k_\mu}{c} P(k_{\mu})
\prod_{\nu=1}^{k_\mu-1} \frac{k_{\mu\nu}}{c} P({k_{\mu\nu}})
\times 
\nonumber \\
& &
\left[1 - \prod_{\mu=1}^k\left(1-\rho\left(1-\prod_{\nu=1}^{k_\mu-1}
\left(1-\rho(1-\tilde h^{k_{\mu\nu}-1})\right)\right)\right)\right],
\label{eq:g2}
\end{eqnarray}

\noindent
and so on. 
Following the reasoning of 
Sec. \ref{Sec:JointGC} 
we can use these equations to identify a 
string of conditional probabilities of infection 
of a node, given its degree and the degree 
configurations of its first and second coordination shells, 
and of further configuration shells if the above iterative 
process were indeed continued. 

\begin{figure}[t!]
\begin{center}
\includegraphics[width=0.4\textwidth]{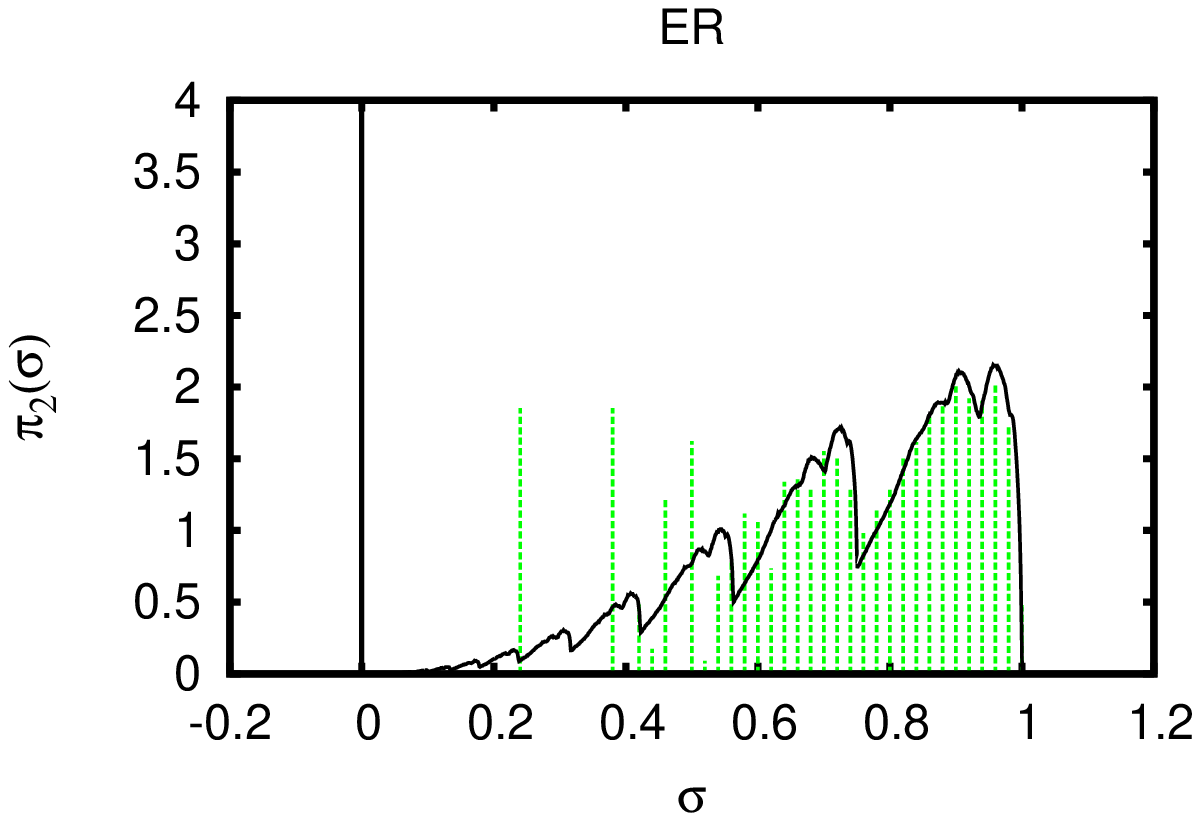}\hfil 
\includegraphics[width=0.4\textwidth]{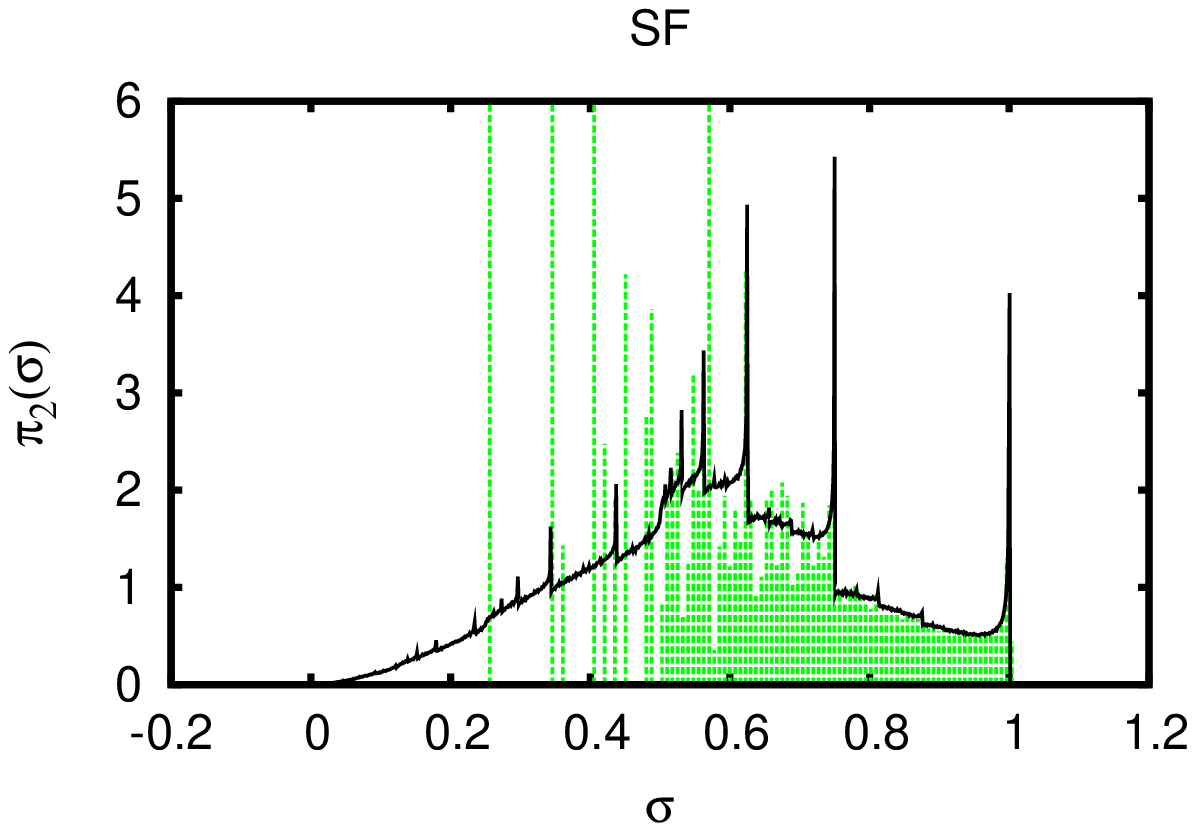} \\
\includegraphics[width=0.4\textwidth]{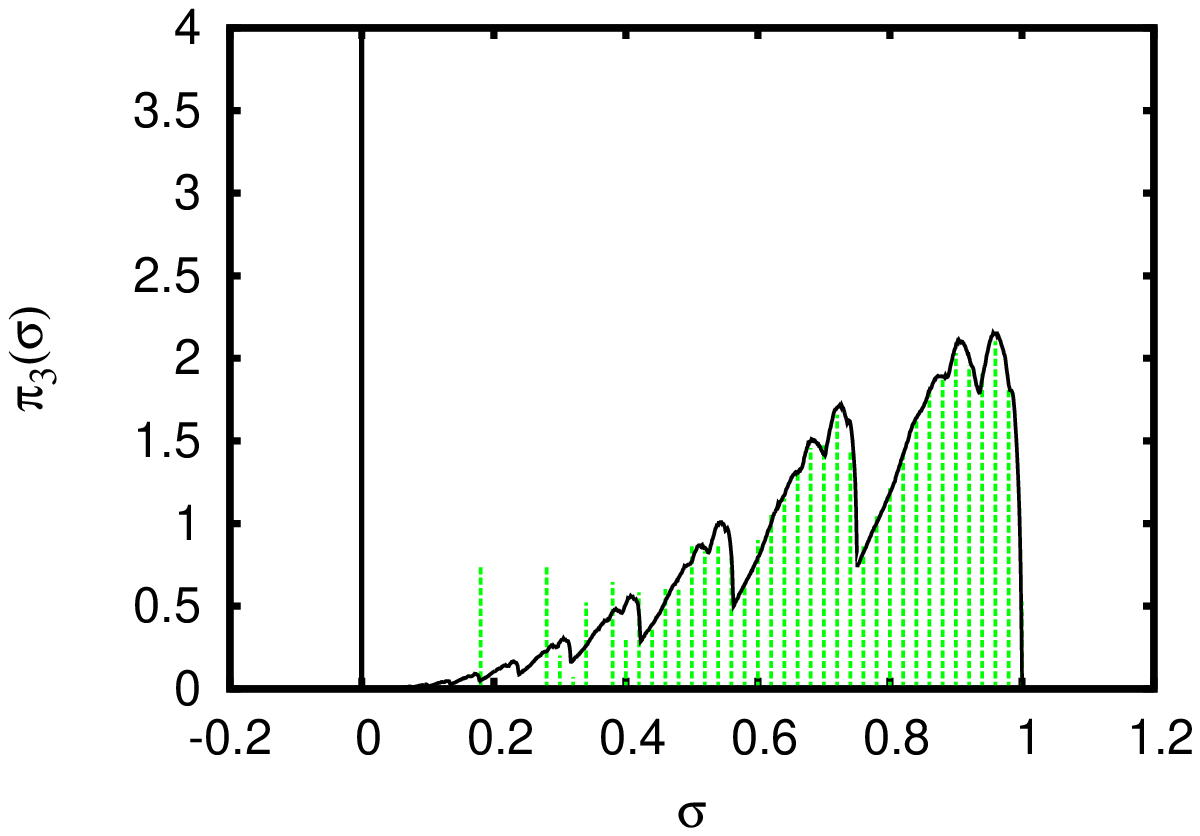}\hfil
\includegraphics[width=0.4\textwidth]{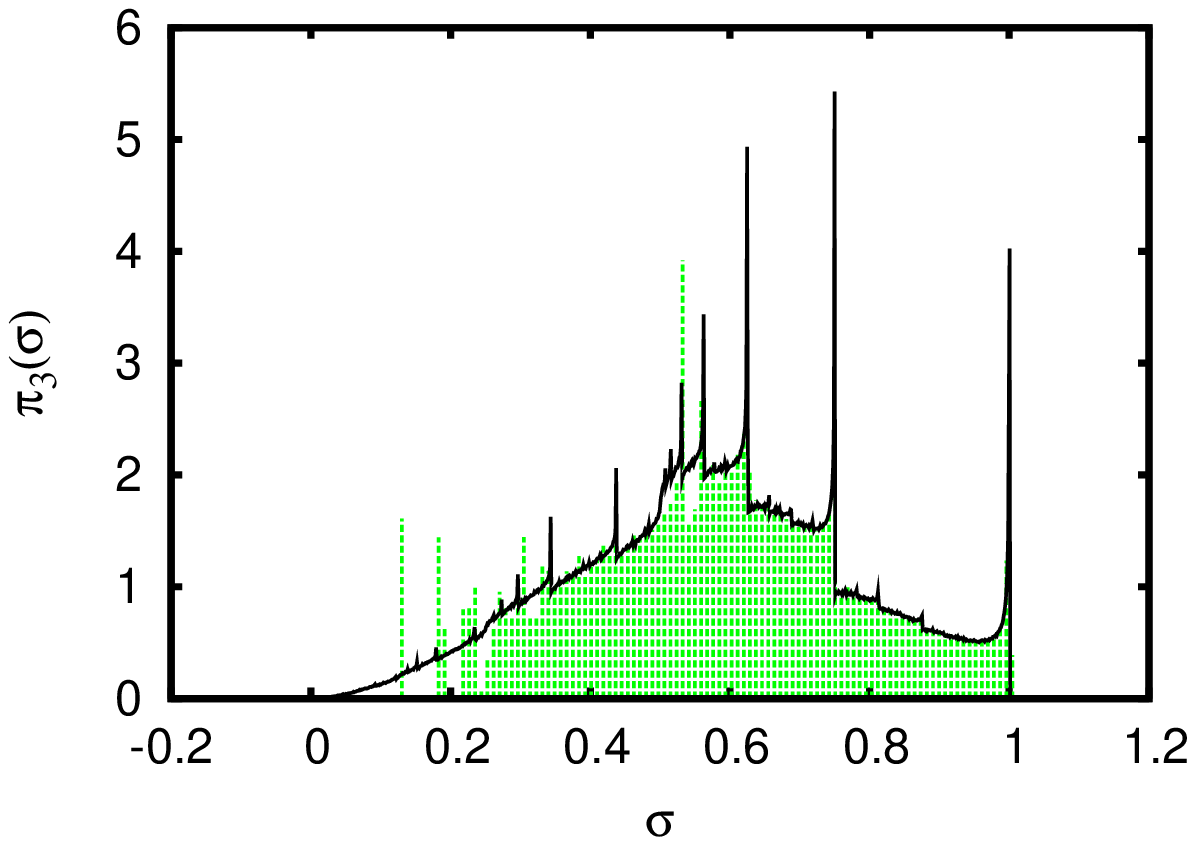} \\
\includegraphics[width=0.4\textwidth]{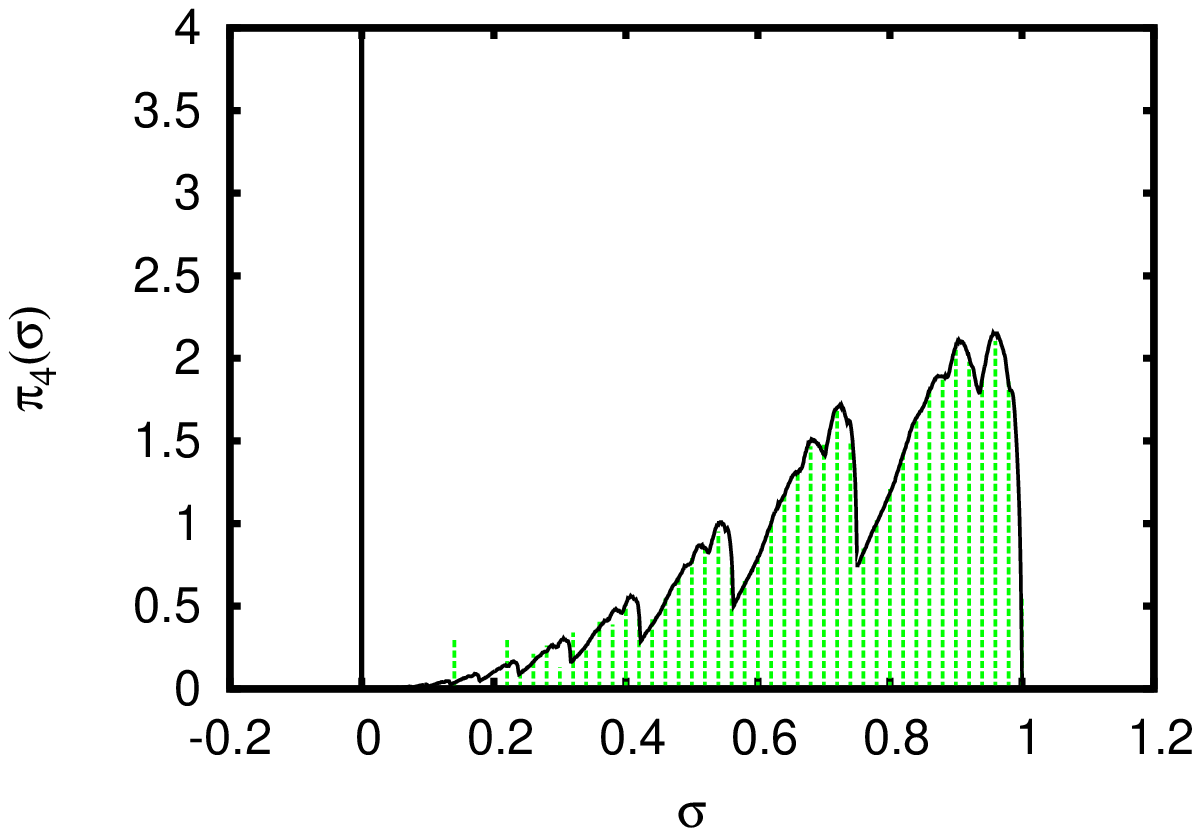}\hfil
\includegraphics[width=0.4\textwidth]{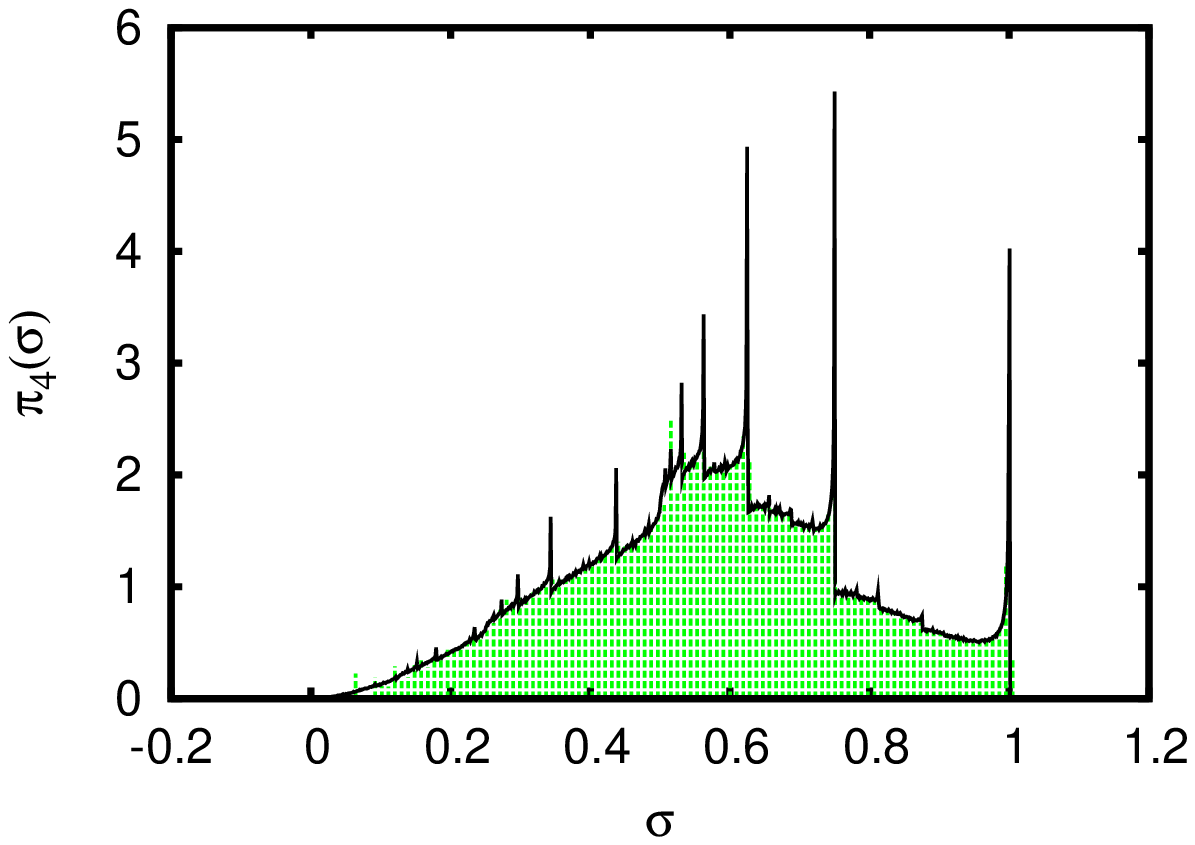}
\end{center}
\caption{
(Color online)
Approximate evaluations of the heterogeneous 
infection probabilities
$\pi_2(\sigma)$, $\pi_3(\sigma)$ 
and $\pi_4(\sigma)$ (top to bottom) represented by green 
dashed histograms, for an ER network with $c=2$ (left column)
and a configuration model network 
(right column)
with a power-law degree
distribution of the form 
$P(k) \sim k^{-\gamma}$ with $\gamma=3$
and $k_{\rm min}=2$, also referred to as a scale-free (SF) network.
Also shown in each panel 
is the full distribution 
$\pi(\sigma)$ (black solid line) obtained using 
a population dynamics approach to solving the self-consistency 
equations arising in this problem. While the low order 
approximations are reasonable at the large $\sigma$ end, higher 
order approximations taking degree configurations of higher 
order coordination shells into account are needed to improve 
accuracy at the small $\sigma$ end of the probability distribution function.
}
\label{fig:13}
\end{figure}

Interestingly, these results allow us to obtain increasingly 
accurate approximations of the full probability density 
function $\pi(\sigma)$ of the heterogeneous infection/percolation 
probabilities defined and evaluated in Ref.
\cite{Kuhn2017}.
In particular, with reference to 
Eqs. (\ref{eq:g0})-(\ref{eq:g2}), 
we define approximations of increasing orders for $\pi(\sigma)$,
starting from the lowest order expression of the form

\begin{equation}
\pi_0(\sigma) = \sum_k P(k) \, \delta \left[ \sigma - (1 - \tilde h^k) \right].
\label{eq:pi0}
\end{equation}

\noindent
Replacing $\tilde h$ in 
Eq. (\ref{eq:pi0}) 
by the right hand side of 
Eq. (\ref{eq:th}),
we obtain  

\begin{equation}
\pi_1(\sigma) = 
\sum_{k;\{k_\mu\}} P(k) 
\prod_{\mu=1}^k \frac{k_\mu}{c} P(k_{\mu}) 
\delta 
\left[ \sigma - \left(1 - \prod_{\mu=1}^k \left(1-\rho(1-\tilde h^{k_\mu-1})\right)\right) \right].
\label{eq:pi1}
\end{equation}

\noindent
Repeating this procedure once more we obtain

\begin{eqnarray}
\pi_2(\sigma) &=& 
\sum_{k;\{k_\mu\};\{k_{\mu\nu}\}}\hspace{-3mm} 
P(k) \prod_{\mu=1}^k
\frac{k_\mu}{c} P(k_{\mu})\prod_{\nu=1}^{k_\mu-1} 
\frac{k_{\mu\nu}}{c} P(k_{\mu\nu})
\times
\nonumber \\
& & \delta
\left[\sigma -
\left(1 - \prod_{\mu=1}^k\left(1-\rho\left(1-\prod_{\nu=1}^{k_\mu-1}
\left(1-\rho(1-\tilde h^{k_{\mu\nu}-1})\right)\right)\right)\right)\right],
\label{eq:pi2}
\end{eqnarray}

\noindent
and so on. 
We note that each of these approximate 
probability density functions of the heterogeneous percolation probabilities 
reproduces the same (exact) average percolation 
probability $\sigma$, as can easily be checked by 
evaluating the averages, and using the iterated  
expressions of Eqs. (\ref{eq:g0})-(\ref{eq:g2}) for $\sigma$.

In Fig. \ref{fig:13} we show evaluations of $\pi_2(\sigma)$, 
$\pi_3(\sigma)$ and $\pi_4(\sigma)$ for an ER network with mean degree of 
$c=2$, with transmission probability 
(bond occupation probability) 
$\rho=0.75$, 
and for a scale free graph with $P(k) \sim k^{-3}$
and $k_{\rm min}=2$, with a transmission probability $\rho=0.5$. 
From the full distribution $\pi(\sigma)$ and in 
particular from its deconvolution according to the degrees 
of the central node, one can conclude that long dangling 
chain configurations (with few side-branches) are mainly 
responsible for the small-$\sigma$ features of $\pi(\sigma)$, and to 
describe sufficiently long chains of this type one would 
have to include higher-order coordination shells in the analysis. 
But the trend towards a reasonably precise description of $\pi(\sigma)$ 
using these low order approximations is clearly visible. Given that 
the present approach is clearly both conceptually and computationally 
`low-tech' compared with the full theory exploited in 
\cite{Kuhn2017}, 
the present low order approximation manage to accurately reproduce a 
considerable amount of detail of these highly non-trivial distributions. 
For the purpose of the evaluation of the approximations, we used sampling 
from randomly generated degree configurations up to the highest order 
configuration shell involved rather than a full evaluation of the sums 
appearing in 
Eqs. (\ref{eq:pi0})-(\ref{eq:pi2}),
as well as the equations for 
$\pi_3(\sigma)$ and $\pi_4(\sigma)$ 
which are not displayed above.

\section{Summary and Discussion}
\label{Sec:Summary}

We have studied the micro-structures of the giant component and the
finite components of configuration model networks.
We found that the finite components form a sub-percolating 
configuration model network with a modified degree distribution.
In particular, the degree distribution of the finite components
is simply an exponentially attenuated version of the original degree
distribution of the network
[Eq. (\ref{eq:pk0e})].
We recovered the known result for the self duality of the ER network
\cite{Molloy1995,Bollobas2001}. 
This self duality means that
the finite components of an ER network with $c>1$ 
form a sub-percolating ER network with mean degree $c_0 = (1-g)c < 1$.
Moreover, we extended this result to a broad class of configuration model
networks. This includes the configuration model with an exponential degree distribution 
and the configuration model with a power-law degree 
distribution and an exponential cutoff.

In contrast, we found that the giant component of a configuration model network
is not itself a configuration model network.
In fact, it exhibits degree-degree correlations to all orders,
as exemplified by our analysis of the percolation problem on the giant component,
the spectrum of the giant component and the DSPL on the giant component.
We presented analytical results for the degree distribution on the giant
component as well as the joint degree distribution of pairs of adjacent nodes.
Furthermore, we provided a methodology for the derivation of the joint
degree distribution of a random node together with several shells around it.
We derived an expression for the assortativity coefficient of the giant component.
Interestingly, the giant component was found to be disassortative, namely high
degree nodes tend to connect preferentially to low degree nodes.
This can be understood intuitively due to the fact that 
disassortativity helps to maintain the integrity of the giant component.
In contrast, the segregation between high degree nodes and low degree nodes
would fragment the giant component into small pieces.
In general, we found that as the network approaches the percolation transition
from above and the giant component decreases in size, its structure becomes more
distinct from the structure of the overall network.
In particular, the degree distribution deviates more strongly from the
overall degree distribution, the degree-degree correlations become stronger
and the assortativity coefficient becomes more negative.

The results presented in this paper have broad implications for dynamical processes
on configuration model networks. For example, epidemic processes are most consequential
when they occur on the giant component. If an epidemic starts on a node which resides on a
finite component it quickly terminates after infecting a small, non-extensive, number of nodes.
In contrast, an epidemic which starts on a node which resides on the giant component
endangers a significant fraction of the entire network. Therefore, the quantities of 
interest in the context of epidemics (as well as many other dynamical processes)
are those that characterize the giant component rather than the overall network.
The examples discussed in this paper demonstrate the difference between the
properties which are conditioned on the giant component and the corresponding 
properties of the entire network.

Our results for the degree distribution and the degree-degree correlations
on the giant component provide a practical and straightforward
way to calculate the properties of many dynamical processes
conditioned on the giant component.
Such processes include information spreading, search processes, network attacks and
random walks. This can be done by utilizing existing theoretical formulations which were
derived for configuration model networks and replacing the degree distribution of the overall
network by the degree distribution conditioned on the giant component.
A more complete formulation can be obtained by incorporating joint
degree distributions, which capture the degree-degree correlations.
The examples discussed in this paper show that 
adapting existing theoretical formulations to account for the special properties
of the giant component provide better approximations than those obtained using the corresponding properties
of the overall network.

\end{document}